\documentclass[pdflatex,sn-mathphys-num]{sn-jnl}% Math and Physical Sciences Numbered Reference Style
\usepackage{graphicx}%
\usepackage{multirow}%
\usepackage{amsmath,amssymb,amsfonts}%
\usepackage{amsthm}%
\usepackage{mathrsfs}%
\usepackage[title]{appendix}%
\usepackage{xcolor}%
\usepackage{textcomp}%
\usepackage{manyfoot}%
\usepackage{booktabs}%
\usepackage{algorithm}%
\usepackage{algorithmicx}%
\usepackage{algpseudocode}%
\usepackage{listings}%
\usepackage{lineno}%
\usepackage{multibib}%
\newcites{A}{References}%
\newcites{B}{Supplementary References}%
\theoremstyle{thmstyleone}%
\theoremstyle{thmstyletwo}%

\theoremstyle{thmstylethree}%

\begin{document}

\title[Article Title]{Anisotropic Dopant and Strain Architectures in WS$_2$ Nanocrystals Driven by Growth Kinetics}


\author*[1]{\fnm{Frederico B.} \sur{Sousa}}\email{fbsousa@ufscar.br}
\equalcont{These authors contributed equally to this work.}

\author[2]{\fnm{Raphaela} \sur{de Oliveira}}
\equalcont{These authors contributed equally to this work.}

\author[3]{\fnm{Matheus J. S.} \sur{Matos}}

\author[4,5]{\fnm{Elizabeth Grace} \sur{Houser}}

\author[3]{\fnm{Igor Ferreira} \sur{Curvelo}}

\author[4,6]{\fnm{Zhuohang} \sur{Yu}}

\author[4,5]{\fnm{Mingzu} \sur{Liu}}

\author[7]{\fnm{Felipe} \sur{Menescal}}

\author[1]{\fnm{Gilmar Eugenio} \sur{Marques}}

\author[7]{\fnm{Leandro M.} \sur{Malard}}

\author[4,5,6]{\fnm{Mauricio} \sur{Terrones}}

\author*[8]{\fnm{Bruno R.} \sur{Carvalho}}\email{brunorc@fisica.ufrn.br}

\author*[7]{\fnm{Helio} \sur{Chacham}}\email{chacham@fisica.ufmg.br}

\author*[1]{\fnm{Marcio D.} \sur{Teodoro}}\email{mdaldin@ufscar.br}

\affil*[1]{\orgdiv{Departamento de Física}, \orgname{Universidade Federal de São Carlos}, \orgaddress{\city{São Carlos}, \state{São Paulo}, \postcode{13565-905}, \country{Brazil}}}

\affil[2]{\orgname{Brazilian Synchrotron Light Laboratory (LNLS), Brazilian Center for Research in Energy and Materials (CNPEM)}, \orgaddress{\city{Campinas}, \state{São Paulo}, \postcode{13083-100}, \country{Brazil}}}

\affil[3]{\orgdiv{Departamento de Física}, \orgname{Universidade Federal de Ouro Preto}, \orgaddress{\city{Ouro Preto}, \state{Minas Gerais}, \postcode{35400-000}, \country{Brazil}}}

\affil[4]{\orgname{Center for 2-Dimensional and Layered Materials, The Pennsylvania State University}, \orgaddress{\city{University Park}, \state{PA}, \postcode{16802}, \country{United States of America}}}

\affil[5]{\orgdiv{Department of Physics}, \orgname{The Pennsylvania State University}, \orgaddress{\city{University Park}, \state{PA}, \postcode{16802}, \country{United States of America}}}

\affil[6]{\orgdiv{Department of Materials Science and Engineering}, \orgname{The Pennsylvania State University}, \orgaddress{\city{University Park}, \state{PA}, \postcode{16802}, \country{United States of America}}}

\affil[7]{\orgdiv{Departamento de Física}, \orgname{Universidade Federal de Minas Gerais}, \orgaddress{\city{Belo Horizonte}, \state{Minas Gerais}, \postcode{30123-970}, \country{Brazil}}}

\affil[8]{\orgdiv{Departamento de Física}, \orgname{Universidade Federal do Rio Grande do Norte}, \orgaddress{\city{Natal}, \state{Rio Grande do Norte}, \postcode{59078-970}, \country{Brazil}}}


\abstract{Dopant distribution in two-dimensional semiconductors is typically assumed to be stochastic, limiting deterministic defect engineering. Here, we show that non-equilibrium growth kinetics can be harnessed to define dopant-driven strain architectures in vanadium-doped WS$_2$ monolayers. Using synchrotron X-ray fluorescence, we identify preferential vanadium incorporation, anti-correlated with tungsten content, along crystallographic bisectors. An adsorption-growth-diffusion model with a single kinetic parameter quantitatively captures the dopant segregation arising from preferential corner adsorption and limited diffusion during chemical vapor deposition growth. Hyperspectral Raman imaging demonstrates mechanically induced vibrational responses, revealing localized tensile strain ($\varepsilon \approx0.70\%$) channels associated with the anisotropic dopant distribution. This regime is marked by the depletion of W-site-sensitive in-plane modes and the emergence of a localized $J2$ mode (210~cm$^{-1}$), which our ab-initio calculations attribute to antiphase V$-$V oscillations. These findings establish kinetic segregation as a route to deterministic chemical and strain architectures in 2D semiconductors, enabling programmable defect landscapes and strain engineering during synthesis.}




\maketitle

\section{Main}

Defect engineering is a powerful route to tailor the electronic, optical, and magnetic properties of two-dimensional (2D) transition metal dichalcogenides (TMDs)~\citeA{lin2016defect,rhodes2019disorder,sousa2025optical}. Substitutional dopants can activate ferromagnetism~\citeA{yun2020ferromagnetic,zhang2020monolayer,pham2020tunable}, modulate carrier polarity~\citeA{kozhakhmetov2021controllable,gao2024electrical}, and reshape excitonic~\citeA{sousa2024effects,mathela2025understanding}, phononic~\citeA{zou2023raman}, and valley-dependent~\citeA{zhou2020synthesis,li2020enhanced,nguyen2021spin,sousa2024giant} responses. These functionalities, however, rely critically on the spatial distribution of impurities. Despite progress in chemical vapor deposition (CVD) growth, the mesoscopic arrangement of dopants in 2D TMDs~\citeA{kumar2025substitutional,jia2025synthesis} remains poorly resolved because existing techniques provide only partial or indirect elemental information. This limited chemical resolution prevents the rational design of defect landscapes and restricts dopant engineering to trial-and-error strategies.

Atomic‐resolution microscopies such as scanning transmission electron microscopy (STEM) and scanning tunneling microscopy (STM) reveal individual substitutional sites~\citeA{zhou2013intrinsic,petHo2018spontaneous,sousa2025optical} but sample only micrometer-scale regions; thus precluding access to long-range spatial patterns. Conversely, optical spectroscopies provide mesoscale coverage~\citeA{sousa2025optical,sousa2025nonlinear} but lack elemental specificity, leading to indirect or conflicting interpretations. As a result, the microscopic origin of the distinct optical anomalies that emerge along the bisector lines of triangular WS$_2$ domains~\citeA{cong2014synthesis,bao2015visualizing,liu2016fluorescence,xu2018microstructure,kastl2019effects,an2020growth,magnozzi2021local,rosa2022investigation} has remained unresolved. Prior reports attributed these features to grain boundaries~\citeA{bao2015visualizing,liu2016fluorescence,xu2018microstructure,kastl2019effects}, vacancies~\citeA{cong2014synthesis,an2020growth}, or surface adsorbates~\citeA{rosa2022investigation}, but a definitive chemical identification has remained elusive.

Here, we overcome this limitation by using synchrotron X-ray fluorescence (XRF) microscopy to directly map the chemical composition of vanadium-doped WS$_2$ monolayers with high sensitivity and mesoscale coverage. The XRF maps reveal a strongly anisotropic dopant landscape: vanadium accumulates along the crystallographic bisectors, forming dopant segregation channels that are sharply anti-correlated with tungsten content. The sulfur sublattice, by contrast, exhibits a spatially homogeneous higher density of sulfur vacancies. These results show that widely used probes have limited capability to resolve spatial heterogeneities of defect and dopant populations in 2D materials, and establish synchrotron nanoprobes as a direct, quantitative metrology for chemical mapping.

To explain the origin of these segregation channels, we develop a continuum adsorption–growth–diffusion model that captures the interplay between preferential corner incorporation and thermal diffusion during growth. This model quantitatively reproduces the measured concentration profiles using a single kinetic parameter, demonstrating that dopant incorporation during CVD growth is dictated by deterministic kinetic pathways rather than random alloying. The mechanism may apply to other dopants and TMD systems and geometries.

Finally, we show that these dopant-rich channels also produce a distinct structural and vibrational response. Hyperspectral Raman imaging reveals a pronounced softening of the $E^{\prime}$ phonon mode along the bisectors, inconsistent with mass or chemical-pressure effects. By disentangling strain and doping contributions, we attribute this behavior to a localized tensile strain field arising from a subtle topographic uplift at the segregation channels. This regime is marked by two spectroscopic signatures: a depletion of the host in-plane $E^{\prime}$ mode and the emergence of a localized $J2$ mode at 210~cm$^{-1}$. Density functional theory (DFT) calculations assign the $J2$ mode to an antiphase in-plane oscillation of neighboring V atoms, indicating the formation of a locally alloyed vibrational environment within the channels. Together, these observations indicate that the vibrational response within the bisector channels is governed by structural relaxation associated with dopant segregation, rather than by purely electronic or mass effects.

Our findings resolve long-standing ambiguity surrounding bisector-line anomalies in WS$_2$, demonstrate that dopant incorporation during CVD synthesis is inherently anisotropic, and establish synchrotron nanoprobes as a powerful approach for chemically and structurally profiling dopant landscapes in 2D semiconductors.

\section{Results}
\subsection{Chemical mapping of anisotropic dopant incorporation}
\begin{figure}[htb!]
 \centering
 \includegraphics[width=0.65\linewidth]{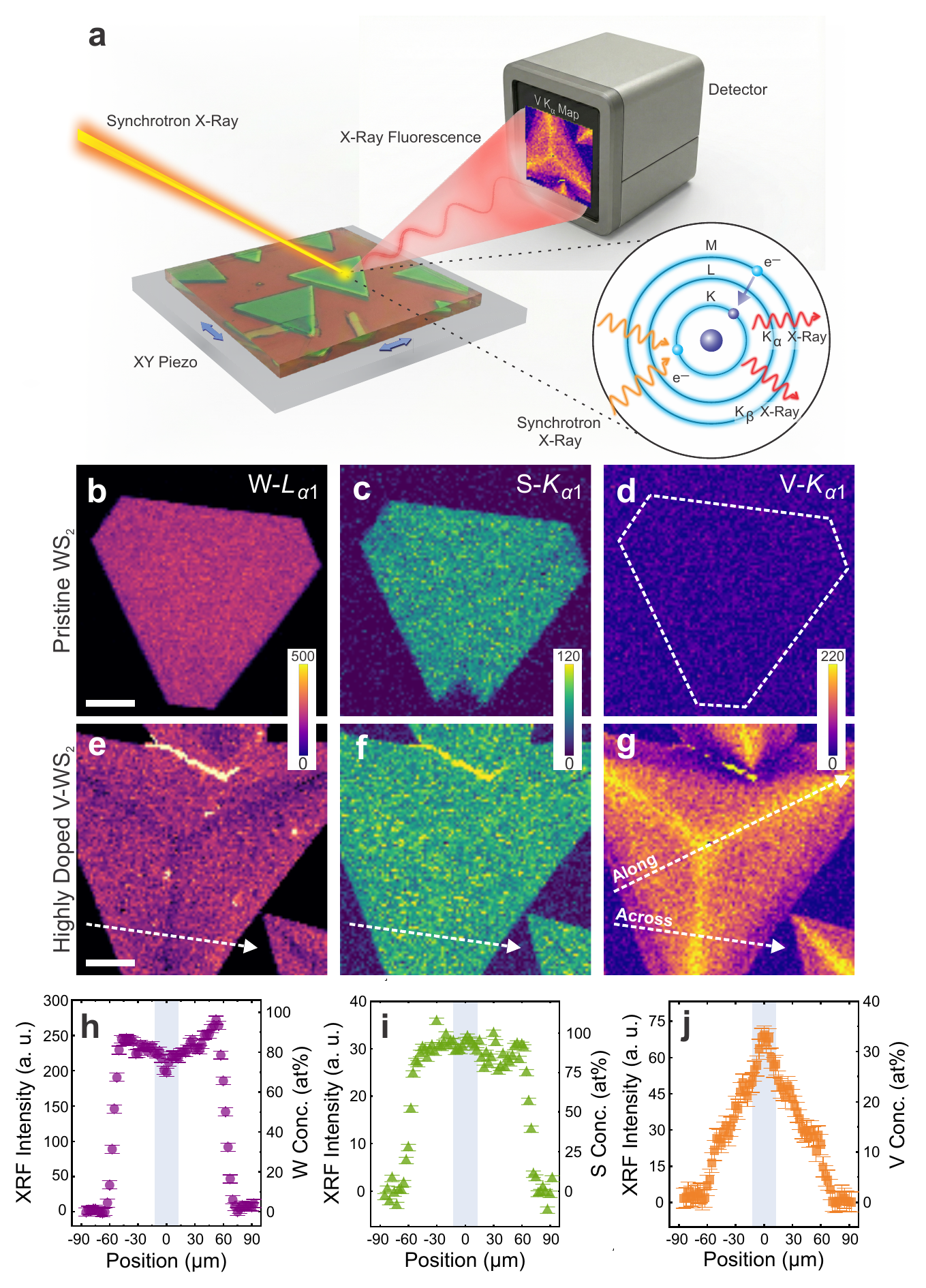} 
 \caption{{\small {\bf Synchrotron X-ray nanoprobe mapping of vanadium segregation.}
    \textbf{a,} Schematic of the CARNAUBA synchrotron X-ray nanoprobe setup. A nanofocused beam excites the sample; the inset (bottom right) illustrates the core-shell ionization and fluorescence emission mechanism (e.g., $K_{\alpha}$, $K_{\beta}$).  
    \textbf{b-g,} XRF intensity maps of pristine ({\bf b-d}) and highly V-doped ({\bf e-g}) triangular domains for tungsten (W-$L_{\alpha}$), sulfur (S-$K_{\alpha}$), and vanadium (V-$K_{\alpha}$). Pristine samples show a uniform host lattice ({\bf d}), whereas doped samples exhibit vanadium accumulation along bisector lines ({\bf g}) that is anti-correlated with tungsten depletion ({\bf e}). Scale bars: 12~$\mu$m ({\bf b}), 50~$\mu$m ({\bf e}).
    \textbf{h-j,} Quantitative concentration profiles across the bisector lines (dashed arrows in ({\bf e-g})). Ratiometric analysis reveals a substitutional gradient peaking at (33$\pm$2)\% V ({\bf j}), with corresponding W depletion to $\sim$70$\pm$1\% ({\bf h}), while the S sublattice remains uniform ({\bf i}). The shaded region indicates the bisector domain.
 }}
 \label{fig_XRF}
 \end{figure}

The experimental configuration for XRF microscopy is shown in Fig.~\ref{fig_XRF}a, where absorption of an incident X-ray photon ionizes an atom, and the subsequent relaxation of higher-energy electrons into the core hole leads to the emission of lower-energy X-ray photons. Because these transitions yield element-specific characteristic energies, XRF provides high chemical sensitivity. We used a fourth-generation synchrotron X-ray nanoprobe (CARNAUBA beamline; see Methods) to raster-scan the samples with a nanofocused beam and detect the characteristic fluorescence emissions from tungsten ($L_{\alpha}$), sulfur ($K_{\alpha}$), and vanadium ($K_{\alpha}$). Representative spectra and integration windows are shown in Supplementary Note~1, Fig.~S1. This approach enables element-specific mapping of large monolayer domains with sub-micrometer resolution~\citeA{mino2018materials,tolentino2023carnauba,winarski2012hard,quinn2021hard,johansson2021nanomax,leake2019nanodiffraction,paterson2011x}, overcoming the field-of-view and sensitivity limits of conventional microfocus X-ray sources.

Pristine WS$_2$ monolayers display a uniform tungsten (W) and sulfur (S) distribution with negligible vanadium (V) signal (Fig.~\ref{fig_XRF}b-d), confirming the chemical homogeneity of the host lattice. In contrast, highly V-doped monolayers exhibit a striking anisotropic chemical texture (Fig.~\ref{fig_XRF}e-g). The V map reveals dopant segregation channels that align with the crystallographic bisector lines, connecting the domain center to the vertices (Fig.~\ref{fig_XRF}g). This enrichment is anti-correlated with local W depletion (Fig.~\ref{fig_XRF}e), whereas the S distribution remains spatially uniform (Fig.~\ref{fig_XRF}f). The inverse correlation between W and V, together with the preserved chalcogen distribution, provides direct chemical evidence for substitutional incorporation of vanadium at tungsten sites ($V_{W}$), rather than interstitial or adsorbed configurations~\citeA{an2020growth,rosa2022investigation}.

To quantify this chemical landscape, we developed a calibration framework that converts XRF intensities into absolute atomic concentrations (Fig.~\ref{fig_XRF}h-j). A stoichiometric baseline was extracted from the pristine sample, which yielded a tungsten reference intensity $I_{W,ref} = 283 \pm 1$ counts. The local W concentration at any point $p$ in the doped monolayer is then:

\begin{equation}
    C^{W}(p) = 100 \times \frac{I_{W}(p)}{I_{W,ref}}
    \label{eqCW}
\end{equation}

The vanadium concentration was determined (see Methods and Supplementary Note~2, Fig.~S2) by analyzing the intensity ratio between points along the bisector channels ($p_1$) and off-bisector regions ($p_2$). Assuming strictly substitutional incorporation, $C^W = 100 - C^V$, the local V concentration is given by:

\begin{equation}
    C^{V}(p_1) = 100 \times \frac{I_{V}(p_1) \left[I_{W}(p_1)-I_{W}(p_2) \right]}{I_{V}(p_2)I_{W}(p_1) - I_{V}(p_1)I_{W}(p_2)},
    \label{eqCV}
\end{equation}

Applying these two estimation frameworks to the XRF intensity profiles across a bisector line (Fig.~\ref{fig_XRF}h-j), a pronounced compositional gradient is revealed. At the segregation channel, the W concentration drops to (70$\pm$1)\%, corresponding to a V incorporation of (33$\pm$2)\%. Toward the flake edges, the W content recovers to (91$\pm$1)\%, with (12$\pm$2)\% V. These measurements unambiguously identify the bisector lines as preferential dopant incorporation pathways during synthesis. Similar anisotropic segregation patterns were observed (Supplementary Note~3, Fig.~S3) for additional XRF scans in pristine and highly doped flakes, as well as for a moderately doped monolayer, where V concentrations of $\sim$7\% at the bisectors and $\sim$3\% at the edges were clearly resolved. These results highlight the high sensitivity of synchrotron XRF nanoprobes for detecting dilute and spatially inhomogeneous dopant populations in 2D materials.

The calibration analysis (Eq.~1) was also employed to estimate a relative sulfur concentration of (89$\pm$2)\% in the highly doped sample, using the average S-$K_{\alpha}$ intensity of 33$\pm$2 from the pristine monolayer as reference. This significantly higher sulfur-vacancy density is in accordance with previous reports~\citeA{zhang2020monolayer,mathela2025understanding}. Moreover, since sulfur vacancies mediate room-temperature ferromagnetism in V-doped TMDs~\citeA{younas2025room}, this elevated vacancy population is an important feature of the doped lattice.

\subsection{Growth-kinetic origin of dopant segregation}
\begin{figure}[htb!]
 \centering
 \includegraphics[width=0.8\linewidth]{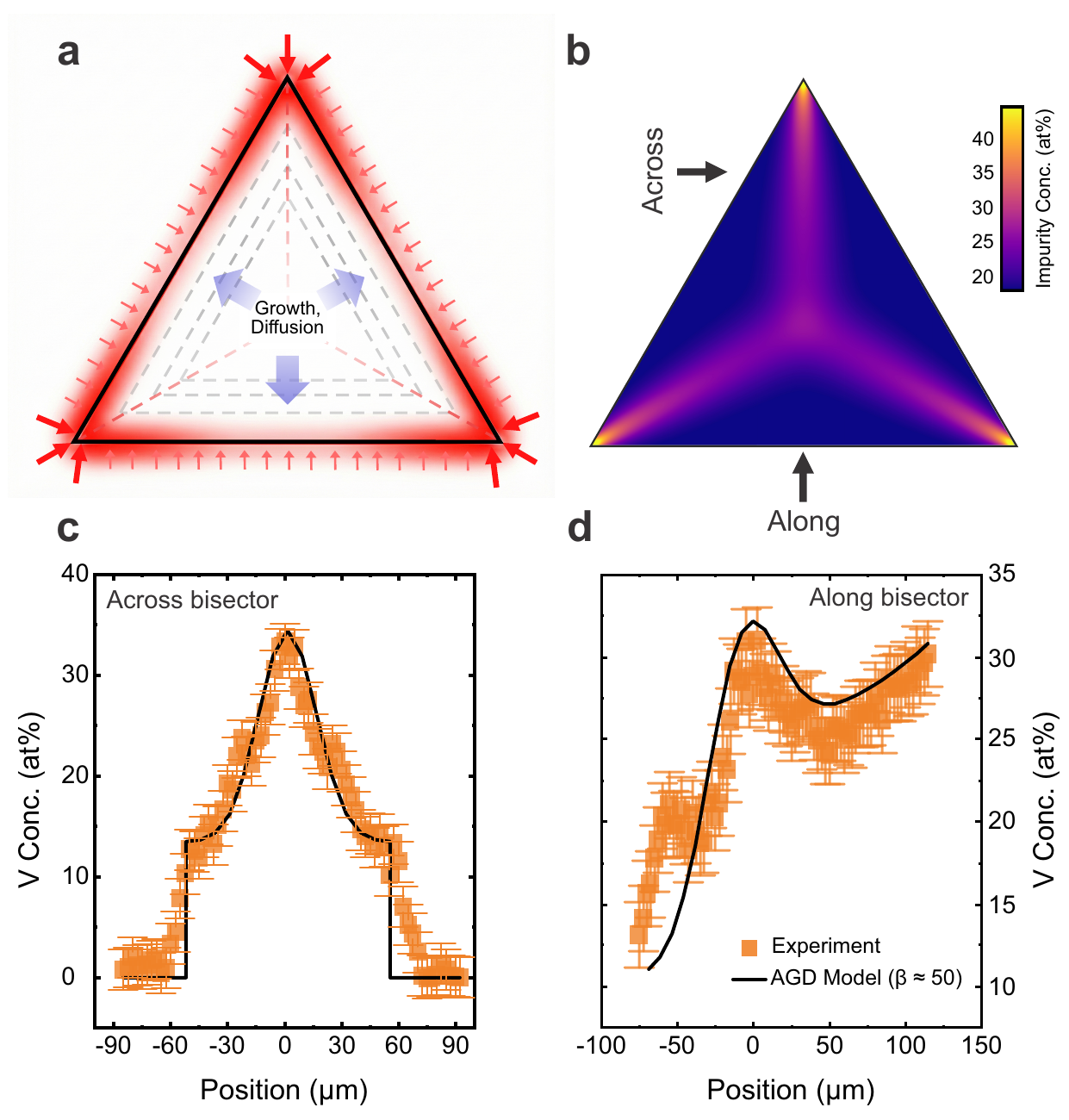} 
 \caption{{\small {\bf Thermodynamic and kinetic mechanism of anisotropic segregation.}
    \textbf{a,} Schematic description of the Adsorption-Growth-Diffusion (AGD) model. Impurity incorporation is governed by two competing pathways: homogeneous adsorption at edges (thin red arrows) and preferential adsorption at high-energy corner sinks (thick red arrows). As the crystal expands (blue arrows), thermal diffusion drives dopant redistribution away from the ideal bisector trajectories (red dashed lines).
    \textbf{b,} Simulated 2D vanadium concentration map generated using the fitted kinetic parameter $\beta = {vL}/{4D} \approx 50$. The model reproduces the characteristic segregation channel structure observed experimentally. 
    \textbf{c,d,} Experimental vanadium profiles extracted across ({\bf c}) and along ({\bf d}) the bisector lines, corresponding to dashed arrows in ({\bf b}). Solid lines represent the analytical AGD model fits. The model accurately captures the sharp accumulation channel ({\bf c}) and the non-monotonic gradient along the bisector ({\bf d}), which arises from the constructive superposition of diffusion fields near the domain center.
 }}
 \label{fig_model}
 \end{figure}

To elucidate the mechanisms underlying this segregation, we developed a continuum analytical model capturing the competition between thermodynamic driving forces and kinetic constraints during crystal growth. In principle, configurational entropy favors a spatially random dopant distribution~\citeA{vlcek2021thermodynamics}. However, DFT shows that the formation energy of tungsten vacancies is substantially lower at the corners of triangular WS$_2$ flakes by about $\sim$3.9~eV relative to edge sites~\citeA{rosa2022investigation}. Extending this argument to substitutional doping suggests that, as the domain grows by atom adsorption, corners behave as preferential incorporation sinks. As growth proceeds outward, the three corner-to-center trajectories (i.e., the bisector lines) accumulate a higher instantaneous defect concentration (Fig.~\ref{fig_model}a). This energetic analysis is consistent with recent reports of anisotropic defect distribution dependent on the edge termination in hexagonal TMD monolayers~\citeA{zhang2023spatial}.

To formalize this mechanism, we introduce an adsorption-growth-diffusion (AGD) model (Fig.~\ref{fig_model}a). We consider a triangular crystal expanding from a nucleation seed to a final center-to-corner distance $L$ at velocity $v$. During growth, impurity incorporation occurs via two distinct pathways: (i) homogeneous adsorption along edges, generating a static maximum-entropy background level $B$, and (ii) enhanced adsorption at energetically favorable corners, producing a high-density amplitude $A$. In the absence of diffusion, corner-incorporated impurities would align into perfectly straight trajectories along the bisectors (red dashed lines in Fig.~\ref{fig_model}a). However, thermal fluctuations drive dopant redistribution, away from the ideal defect lines, into the domain interior according to Fick’s second law. Solving the time-dependent diffusion equation for this evolving boundary condition (derivation in Methods and Supplementary Note~4) yields the steady-state concentration profile:

\begin{equation}
    C(r) = A {\mathcal F}_{\beta}(r) + B,
\end{equation}

\noindent where ${\mathcal F}_{\beta} (r)$ is a spatial diffusion function governed by the dimensionless parameter $\beta = {vL}/{4D}$, which captures the balance between the domain expansion rate $v$ and thermal diffusion efficiency $D$. Large $\beta$ signifies fast growth or slow diffusion, effectively ``freezing in'' dopant distribution near its incorporation sites.

Fitting the AGD model to the experimental profiles yields $\beta = 50$ for both across (Fig.~\ref{fig_model}c) and along (Fig.~\ref{fig_model}d) the segregation channels, indicating that dopant transport is strongly kinetically limited during the later stages of growth. 
The transverse profile (Fig.~\ref{fig_model}c) exhibits a Gaussian-like decay, characteristic of diffusion from a line-like dopant source. The longitudinal profile (Fig.~\ref{fig_model}d), however, displays a non-monotonic concentration variation arising from two competing mechanisms captured by the AGD model. Near the domain center ($d = 0~\mu \text{m}$), constructive overlap of the three diffusion fields emerging from the converging bisectors produces a local maximum. As the bisectors diverge, the overlap weakens, producing a dip concentration. At larger distances ($d > 50~\mu \text{m}$), the concentration rises again due to a kinetic lag: impurities incorporated near the flake tips at late growth stages have considerably shorter diffusion times, preserving a high-density profile.

These results show that dopant incorporation in CVD-grown WS$_2$ is governed by the interplay between preferential corner adsorption and finite-time diffusion, rather than equilibrium thermodynamics alone. The AGD model provides a mechanistic description of the anisotropic segregation channels observed experimentally and may be generalizable to other dopants, layered material systems, and flake geometries.

\subsection{Dopant- and strain-induced vibrational renormalization}
\begin{figure}[htb!]
 \centering
 \includegraphics[width=0.90\linewidth]{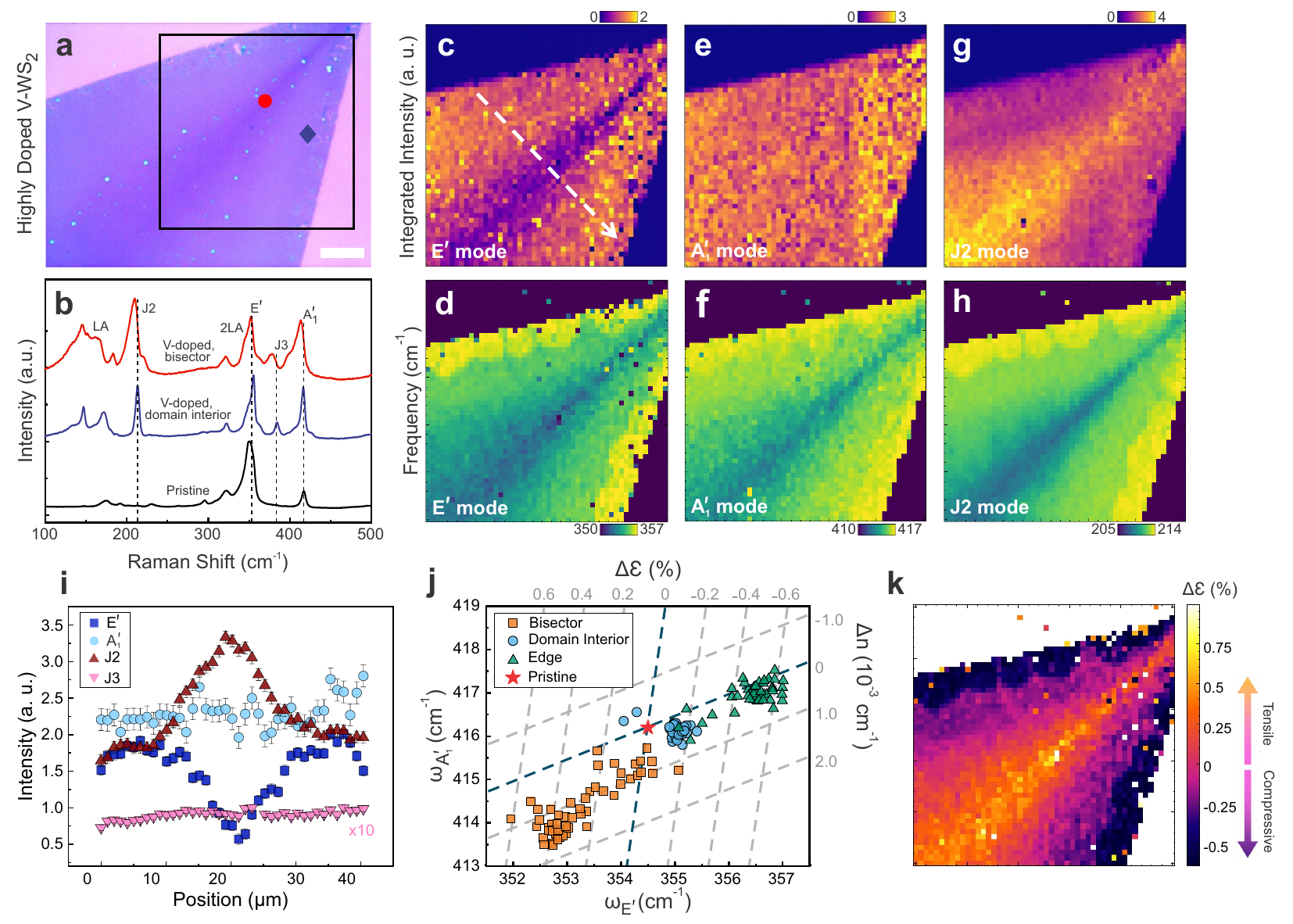} 
 \caption{{\small {\bf Experimental evidence of localized alloy formation and strain patterning at segregation channels.}
    \textbf{a,} Optical image of a highly V-doped WS$_2$ triangular monolayer. The black square denotes the hyperspectral mapping area. Scale bar, 10~$\mu$m.
    \textbf{b,} Representative Raman spectra from the domain interior (blue) and bisector (red) indicated by colored symbols in ({\bf a}). The spectra reveal the emergence of localized $J2$ and $J3$ modes, accompanied by a pronounced redshift of the host $E^{\prime}$ and $A^{\prime}_1$ modes. 
    \textbf{c-h,} Hyperspectral Raman maps of the absolute integrated intensity (top row) and frequency shift (bottom row) for the $E^{\prime}$ ({\bf c, d}), $A^{\prime}_1$ ({\bf e, f}), and $J2$ ({\bf g, h}) modes normalized by silicon peak. While the $E^{\prime}$ intensity ({\bf c}) decays at the bisectors, the $A^{\prime}_1$ intensity ({\bf e}) remains strikingly homogeneous across the flake. The localized $J2$ mode ({\bf g}) peaks sharply at the segregation channels, signaling the emergence of the alloy phase. Scale bar, 10~$\mu$m.
    \textbf{i,} Absolute Raman intensity profiles extracted across the bisector, corresponding to dashed arrow in ({\bf c}). The crossing of the decaying $E^{\prime}$ (blue) and rising $J2$ (dark red) intensities identifies the transition to a localized alloy, whereas $A^{\prime}_1$ (light blue) and $J3$ ($\times$10, magenta) show stable intensity. Error bars denote the spatial standard deviation derived from a 10-point lateral average, reflecting the structural homogeneity of the segregation channel.    
    \textbf{j,} Frequency correlation plot ($\omega_{A_1^\prime}$ vs. $\omega_{E^\prime}$) used to disentangle strain ($\Delta \epsilon$) and doping ($\Delta n$) contributions across the entire flake. While the domain interior (blue) and edges (green) track towards the compressive quadrant, the bisector (orange) aligns exclusively with the tensile mechanical signature (slope $\approx 0.90$). The pristine reference point (red star) serves as the anchor for zero strain/doping.
    \textbf{k,} Spatial map of the relative strain variation derived from the vector decomposition analysis, revealing a localized tensile field of $\varepsilon \approx (0.70 \pm 0.05)\%$ concentrated along the dopant segregation channels. 
 }}
 \label{fig_Raman}
 \end{figure}

Having identified the mechanistic origin of dopant segregation, we examine its impact on the vibrational response of the monolayer. Hyperspectral Raman imaging was performed on the highly doped triangular domain shown in Fig.~\ref{fig_Raman}a, with representative spectra collected at the bisector line (red) and domain interior (blue) (Fig.~\ref{fig_Raman}b).
The pristine WS$_2$ monolayer displays the 2LA (350~cm$^{-1}$), $E^{\prime}$ (355~cm$^{-1}$), and $A_1^{\prime}$ (416~cm$^{-1}$) modes~\citeA{carvalho2020resonance}. In the doped sample, symmetry breaking activates the defect-mediated LA band (175~cm$^{-1}$) and introduces the $J2$ (210~cm$^{-1}$), and $J3$ (380~cm$^{-1}$) modes. While these features have been previously attributed to a vanadium-doping-induced 2H to 1T phase transition~\citeA{han2021one,gao2024electrical}, our stability analysis (see Supplementary Note~5) confirms that the 2H phase remains the thermodynamic ground state. This is consistent with Ref.~\citeA{zou2023raman}, which identifies the $J2$-like mode as a spectroscopic hallmark of heavily V-doped WS$_2$ monolayers within the 2H lattice symmetry.

%As discussed in the following section, these modes emerge as spectroscopic fingerprints of localized V$-$V$_S$ complexes within the 2H framework, arising from local symmetry lowering and structural disorder.

Mapping the spatial variation of these modes reveals pronounced vibrational phenomena localized at the segregation channels (Fig.~\ref{fig_Raman}c-h), while the pristine monolayers are spectrally uniform (Supplementary Note~6, Figs.~S6 and~S7). A fundamental observation is the mode-selective integrated intensity modulation across the segregation channels (Fig.~\ref{fig_Raman}i). The $E^{\prime}$ mode exhibits a localized collapse at the bisector center (Fig.~\ref{fig_Raman}c,i), whereas the $A_1^{\prime}$ area remains homogeneous across the entire flake (Fig.~\ref{fig_Raman}e,i). This contrast reflects the distinct atomic character of these phonons: the in-plane $E^{\prime}$ mode, involving substantial displacement of the metal sublattice, is highly sensitive to the substitution of W by V. By contrast, the out-of-plane $A_1^{\prime}$ mode is dominated by sulfur vibrations; consequently, its vibrational coherence remains comparatively less perturbed by cation substitution. This mode-selective response suggests that the $E^{\prime}$ signal predominantly originates from localized WS$_2$-rich islands, whereas the $A_1^{\prime}$ mode remains active across both pristine and alloyed domains. Particularly, this interpretation is supported by the comparison between WS$_2$ and VS$_2$ monolayers~\citeA{su2020sub,zhang2022effect}, whose Raman spectra exhibit a markedly larger frequency renormalization for the in-plane $E^{\prime}$ mode than for the out-of-plane $A_1^{\prime}$ mode. Moreover, the integrated intensity modulation of the bulk vibrational modes shows a direct correlation with the emergent $J2$ and $J3$ modes (Fig.~\ref{fig_Raman}i). While $J2$ peaks sharply at the bisector, exhibiting an intensity inversion with the host $E^{\prime}$ mode, the $J3$ remains spatially stable, similarly to $A_1^{\prime}$. 
%This connected Raman response indicates that $J2$ might be related to in-plane V vibrations, which is supported by calculations discussed in the following section.
% , while $J3$ serves as a consistent structural baseline

%Notably, despite the reduced atomic mass of vanadium, these impurity-activated modes appear at lower frequencies than the host phonons, reflecting a substantial softening of the local force constants that overrides the expected mass-induced hardening within the alloyed regions.

Phonon linewidths analysis further reveals a mode-selective disorder (Supplementary Note~6, Fig.~S7): the out-of-plane $A_1^{\prime}$ mode broadens significantly, whereas the in-plane $E^{\prime}$ linewidth remains largely unperturbed. Although controversial, this mode-selective response also points to a substitutional disorder and alloying within the metal sublattice, while indicating a minor perturbation of the sulfur sublattice. As discussed in the integrated intensity analysis, the $E^{\prime}$ response comes mostly from localized WS$_2$-like domains, resulting in the conservation of the narrow linewidth despite the intensity quenching. The $A_1^{\prime}$ signal, however, convolutes the out-of-plane vibrations from the disordered alloyed regions, leading to a spectral weight redistribution. Notably, while the amplitude of the $A_1^{\prime}$ mode decreases at the bisector (Supplementary Note~6, Fig.~S8), its concurrent broadening results in the nearly homogeneous integrated area across the flake (Fig.~\ref{fig_Raman}e,i). 

In principle, substitution of W by lighter V should induce phonon hardening due to the reduced mass, and the smaller ionic radius of V should impose local compressive chemical pressure via bond-length contraction, both yielding a blueshift primarily in the in-plane $E^{\prime}$ mode. While the domain interiors and edges exhibit an $E^{\prime}$ hardening (Fig.~\ref{fig_Raman}d), along the bisector lines this trend is completely reversed. Despite having the highest V concentration ($\sim$33\%), both $E^{\prime}$ and $A_1^{\prime}$ modes undergo pronounced softening (redshift). For the $A_1^{\prime}$ mode (Fig.~\ref{fig_Raman}f), this softening reflects the weaker V$-$S bonding relative to W$-$S~\citeA{sevy2017bond,johnson2016predissociation}, which lowers the out-of-plane force constant. Additionally, $A_1^{\prime}$ mostly involves sulfur atoms' vibrations; consequently, the lighter V mass compared to W weakly contributes to the phonon renormalization. Particularly, this $A_1^{\prime}$ redshift is indeed observed in the comparison between WS$_2$ and VS$_2$~\citeA{su2020sub,zhang2022effect} monolayers, where VS$_2$ constitutes the limiting case of V substitution. The large redshift of the $E^{\prime}$ mode ($\sim$3~cm$^{-1}$ relative to the domain edge), however, contrasts with the blueshift reported for the limiting VS$_2$ case~\citeA{su2020sub,zhang2022effect}. Therefore, this strong internal variation, measured within a single continuous domain, might reflect a particular electronic and structural environment at the disordered alloyed regime.

To disentangle the contributions of mechanical strain ($\Delta\varepsilon$) and electronic doping ($\Delta n$), we performed a correlation analysis~\citeA{lee2023suppression} of the $E^{\prime}$ and $A_1^{\prime}$ frequencies (Fig.~\ref{fig_Raman}j). The resulting distribution reveals that while the domain interior and edges indeed track along a compressive strain, the bisector regions, however, deviate sharply from the established doping trajectory for WS$_2$, instead aligning exclusively with a tensile-strain manifold ($\partial \omega_{A_1^{\prime}} / \partial \omega_{E^{\prime}} \approx 0.90\pm0.07$ with $R^{2} > 0.90$). This alignment indicates that the $E^{\prime}$ shift is governed predominantly by a mechanically driven response, overriding the contribution from $p$-type doping introduced by V atoms~\citeA{zhang2020monolayer}. If the measured $\sim$3~cm$^{-1}$ redshift at the bisector were purely electronic, it would require a carrier density variation orders of magnitude higher than our spectroscopic baseline allows. Thus, the correlation analysis confirms that the $E^{\prime}$ softening is driven by a localized tensile strain field of $\varepsilon \approx (0.70 \pm 0.05)\%$, as mapped in Fig.~\ref{fig_Raman}k.

Topographic analysis (Supplementary Note~7, Figs.~S9 and~S10) identifies subtle morphological uplift or ``wrinkling'' along the bisectors, providing a structural origin for this anomalous local tension. The correlation between $E^{\prime}$ and $A_1^{\prime}$ modes was also analyzed under systematically varied isotropic biaxial strain through DFT calculations, showing a quantitative consistency with our experimental results (Supplementary Note~8, Fig.~S11). Crucially, the threefold periodic arrangement of these tensile channels defines a deterministic strain patterning, providing a pathway for exciton funneling and quantum emitter localization~\citeA{parto2021defect,linhart2019localized,fonseca2025observation,liberal2026quantum}.

\subsection{Microscopic origin of lattice disorder}
\begin{figure}[!htb]
    \centering
    \includegraphics[width=0.75\linewidth]{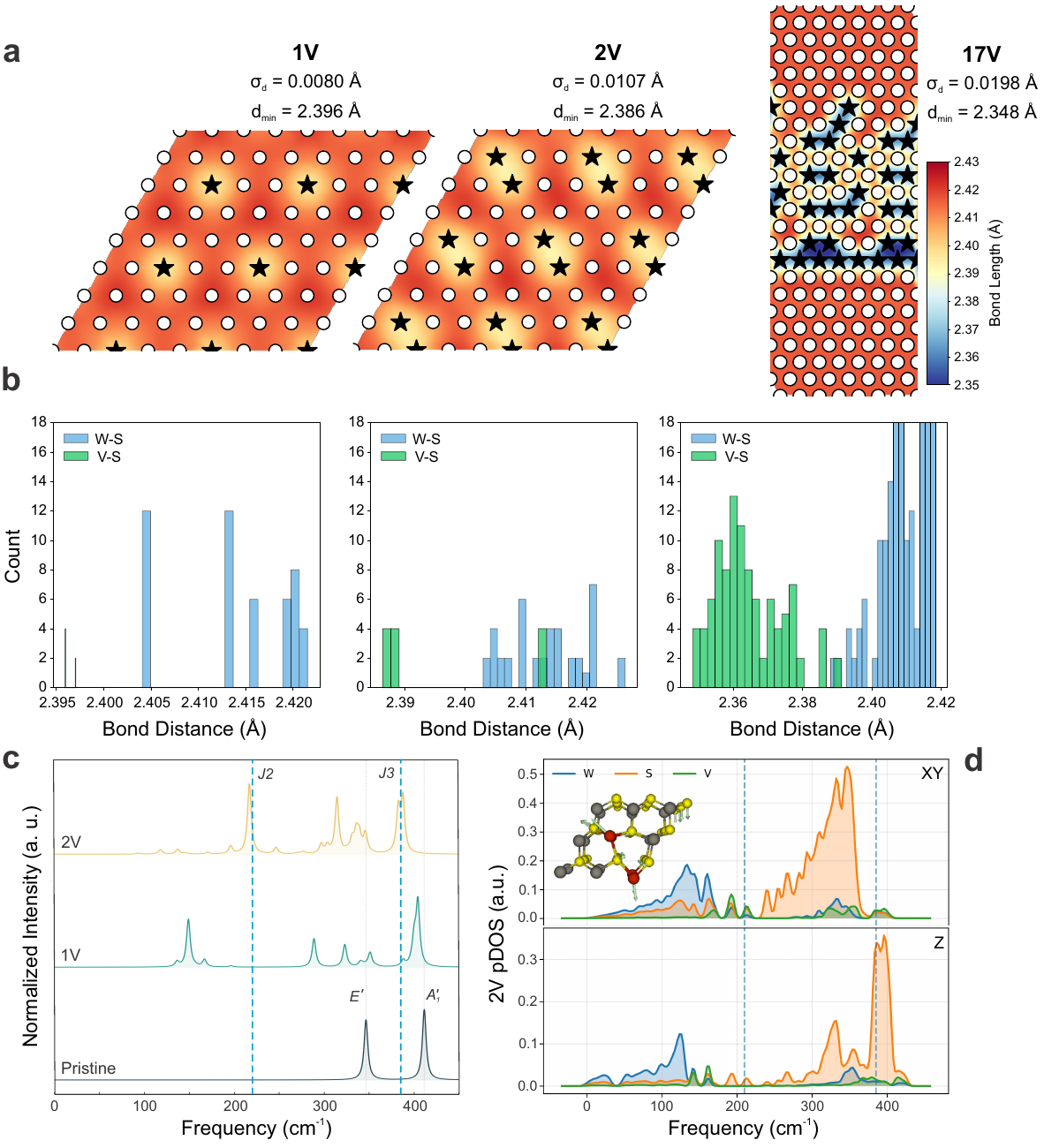}
    \caption{{\bf Ab initio modeling of the localized alloy phase and vibrational signatures.} 
    \textbf{a,} Atomic-scale bond length heatmaps for 1V, 2V, and 17V disordered ribbon (nanowire) models. The spatial distribution of vanadium atoms (stars) within the WS$_2$ host (circles) induces a bipolar distortion landscape, where local coordination shells contract ($\text{d}_\text{min} \approx 2.35$~\AA) while generating mesoscopic tensile relaxation in the surrounding lattice. $\sigma_{d}$ denotes the standard deviation of bond lengths.
    \textbf{b,} Bond distance histograms corresponding to the models in {\bf a}, revealing the emergence of a bimodal distribution in the high-concentration 17V regime. The separation between the V$-$S (green) and W$-$S (blue) populations quantifies the structural disorder of the localized alloy. 
    \textbf{c,} Calculated Raman spectra for pristine, 1V, and 2V supercells. The results show the depletion of the host $E^{\prime}$ and $A_1^{\prime}$ modes and the activation of localized $J$-features (dashed blue). The $J2$ mode emerges as a definitive signature of V$-$V proximity.
    \textbf{d,} Site-projected phonon density of states (pDOS) fir the 2V model, partitioned into in-plane (XY) and out-of-plane (Z) contributions. The $J2$ feature at $\sim210$~cm$^{-1}$ is primarily driven by in-plane metal atom displacements. Inset: vibrational displacement schematic for the $J2$ mode, identifying it as an antiphase in-plane oscillation of neighboring vanadium atoms.
    }
    \label{fig:strainmap}
\end{figure}

The deterministic segregation of vanadium along the crystallographic bisectors of WS$_2$ monolayers gives rise to a complex coupling between mechanically induced strain fields and the local chemical environments. Experimentally, these segregation domains are manifested as characteristic topographic elevations accompanied by pronounced structural wrinkling. To clarify the microscopic mechanisms underlying these observations, we performed DFT calculations to resolve the local structural perturbations arising from V incorporation (Fig.~\ref{fig:strainmap}). 

While the pristine WS$_2$ lattice exhibits spatially homogeneous W$-$S bond lengths (2.41$~$\AA), substitutional V$_W$ defects act as primary centers of lattice distortion. In the dilute limit (1V, Fig.~\ref{fig:strainmap}a), the lattice displays localized bond-expansion regions associated with isolated impurity strain fields. As the dopant concentration increases, these fields progressively overlap, leading to a cooperative amplification of lattice disorder and a systematic increase in the bond-length standard deviation ($\sigma_d$). For the 3V configuration ($\sim$33\%~V), the structural disorder is characterized by $\sigma_d = 0.0125$~\AA~and a minimum bond distance of $d_{min} = 2.383$~\AA.

To bridge the gap between unit-cell models and the mesoscopic heterogeneity observed along the bisectors, we constructed a periodic stripe-patterned WS$_2$ supercell containing a central V-rich region with 17 randomly distributed vanadium atoms, corresponding to a local concentration of approximately 35\%. This geometry represents a periodically repeated V-rich stripe embedded between pristine WS$_2$ domains, thereby mimicking the dopant-rich channels observed experimentally without introducing free edges or finite-size nanoribbon effects (17V, Fig.~\ref{fig:strainmap}a). The corresponding bond-distance histograms (Fig.~\ref{fig:strainmap}b) reveal a striking divergence between the W$-$S and V$-$S populations, where the disordered ribbon exhibits a significant increase in bond-length dispersion ($\sigma_d = 0.0198$~\AA) and a pronounced reduction in the minimum bond distance ($d_{min} = 2.348$~\AA). Notably, the spatial heatmaps reveal a ``bipolar distortion'' pattern where locally compressed V$-$S coordination shells coexist with extended tensile relaxation regions, providing the structural foundation for the $E^{\prime}$ redshift observed in our Raman mapping. Physically, the steep structural gradient across the ribbon facilitates a three-dimensional relaxation of the 2D lattice, manifesting as topographic uplift and localized tensile-strain channels, with a theoretical pristine-WS$_2$ slope of $\partial\omega_{A'_1}/\partial\omega_{E'} \approx 0.98$ (Supplementary Note~S8, Fig.~S11), in accordance with our experimental results.

The impact of this structural environment on the vibrational properties is captured by the simulated Raman spectra (Fig.~\ref{fig:strainmap}c and Supplementary Note~8, Figs.~S12 and~S13). Our calculations show that $A_1^{\prime}$ intensity remains largely unperturbed for the 1V configuration, while the $E^{\prime}$ mode is significantly affected. This agrees with our experimental results (Fig.~\ref{fig_Raman}), confirming the dominant disorder in the metal sublattice, with only minor perturbations to the chalcogen sublattice, leading to a preferential loss of coherence in the in-plane $E^{\prime}$ mode. Specifically, this corroborates the hypothesis that $E^{\prime}$ only survives at WS$_2$-rich islands, whereas $A_1^{\prime}$ remains active at alloyed domains. Additionally, the Raman calculations agree with the experimentally observed frequency softening and linewidth broadening of the $A_1^{\prime}$ mode.

The DFT analysis also evidences the emergence of the $J2$ mode as a definitive signature of high vanadium concentration, appearing prominently as V$-$V proximity increases in the 2V model. The symmetry reduction inherent in this distorted nanowire architecture breaks the hexagonal $D_{3h}$ symmetry, relaxing the selection rules and activating vibrations that are otherwise forbidden in the pristine lattice. Specifically, our calculations identify the $J2$ feature ($\sim$210~cm$^{-1}$) as a localized vibrational mode involving metal-atom motion and sulfur scissoring. As illustrated in the vibrational displacement schematic (Fig.~\ref{fig:strainmap}d, inset, and Supplementary Fig.~S13), the $J2$ mode consists of an antiphase in-plane oscillation of two neighboring substitutional V atoms at the metal sublattice. The phonon density of states (pDOS, Fig.~\ref{fig:strainmap}d and Supplementary Fig.~S12) further supports the assignment of this mode to the coupling between V-induced local vibrations and the surrounding WS$_2$ lattice. This specific vibrational signature explains why $J2$ intensity peaks so sharply at the bisectors and exhibits an inversion in intensity relative to the host in-plane $E^{\prime}$ mode (Fig.~\ref{fig_Raman}c,g,i), identifying it as a spectroscopic proxy for the localized alloy phase. Moreover, the assignment of the $J2$ mode to V–V pairs is supported by a prior study~\citeA{zou2023raman} that observed an increased $J2$ Raman intensity when there is a prevalence of neighboring V atoms resolved in STM measurements.

%The interplay between electronic and mechanical effects is particularly nuanced in these high-concentration regions. While V$_W$ substitution introduces $p-$type character~\citeA{zhang2020monolayer}, the Raman response is governed by the structural manifold. This indicates that the expected electronic hardening is energetically surpassed by the massive lattice relaxation within the 17V ribbon. Specifically, the $\sim$0.70\% tensile strain provides a dominant softening contribution that exceeds the hardening influence of the hole doping associated with substitutional V incorporation. Consequently, the host $E^{\prime}$ and $A_1^{\prime}$ modes serve as primary probes of the mesoscale tensile state and metal-sublattice disorder, while impurity-activated $J$-modes reflect the degree of alloying. This nanowire architecture thus defines a unique multiscale distortion landscape, where atomic configurational disorder and mesoscopic strain patterning converge to dictate the optical and vibrational identity of the bisector channels.

\section{Discussion}
Our results demonstrate that synchrotron X-ray fluorescence nanoprobe mapping provides a powerful metrology for doped 2D materials, bridging the gap between atomic-resolution microscopies and wide-field optical probes. By achieving quantitative chemical mapping of a doped monolayer, we show that optical anomalies at WS$_2$ bisectors originate from symmetry-dictated dopant segregation channels rather than stochastic defects. This finding suggests that disorder in quantum materials may be inherently designable.

The adsorption-growth-diffusion (AGD) framework identifies a single parameter, $\beta$, that governs the interplay between corner-selective adsorption and diffusion. This indicates that chemical heterogeneity can emerge from growth dynamics rather than extrinsic material irregularities. By tuning the growth velocity or modifying the diffusion landscape (e.g., temperature, precursor flux, or substrate interactions), the width, contrast, and periodicity of segregation channels may become controllable degrees of freedom, enabling the deliberate design of chemical textures in large-area monolayers. 
The predictive nature of the AGD framework further suggests that controlled interruption and resumption of vanadium incorporation during synthesis, analogous to the growth of 2D lateral heterostructures~\citeA{sahoo2018one}, could enable deterministic patterning of dopant architectures. The model predicts that such ``pulsed'' growth would fragment continuous segregation channels into discrete quantum-dot-like regions near the domain center and metallic-contact regions along the bisectors (Supplementary Note~9, Fig.~S14). This points to a lithography-free route for the architecture of functional electronic motifs within a single-layer crystal.
More broadly, the AGD framework may also provide insight into the role of diffusion-driven entropy in doped 2D systems~\citeA{miao2025multivalence,wyatt2025order}.

A second emergent feature revealed by these measurements is the strong coupling between dopant concentration, local structural uplift, and tensile strain along the bisector channels. This coupling produces symmetry-locked strain patterning, in which mechanical deformation is confined to these dopant-rich bisectors that act as deterministic strain potentials. Our data suggest that these channels define multiscale distortion landscapes where atomic-scale configurational disorder and mesoscopic strain patterning converge. Such strain funnels provide a route for exciton guidance, quantum-wire formation, and scalable strain engineering in 2D semiconductors~\citeA{hou2025engineering}.

The transition from a doped host to a localized alloy phase along the bisector channels is reflected in the vibrational identity of these channels. Our combined Raman and DFT analysis shows that the vibrational response is governed by a mechanical manifold, in which structural relaxation of the vanadium-rich regions overrides the expected electronic perturbations. While $p-$type substitution introduces an inherent chemical hardening, the resulting $\sim$0.70\% tensile strain provides a dominant softening contribution that determines the in-plane phonon redshift. Accordingly, the host $E^{\prime}$ mode probes the mesoscale tensile state and metal-sublattice disorder, whereas the emergence of impurity-activated $J$-modes, particularly the $J2$ antiphase V$-$V oscillation, serves as a spectroscopic signature of the localized alloy phase.

These observations point to a state of structural perturbation within the bisector segregation channels: the out-of-plane integrity of the sulfur sublattice remains largely preserved even at high doping levels, while the in-plane metal sublattice evolves toward a localized, disordered alloy phase. This mode-selective response identifies these bisector channels as self-assembled quantum-wire-like architectures embedded within the WS$_2$ host, where atomic-scale bond rearrangement coexists with mesoscopic lattice relaxation.  

Ultimately, our findings transform the perception of dopant segregation from an uncontrolled material imperfection into a functional degree of freedom. By linking growth kinetics to mesoscale chemical and mechanical patterning through a quantitative adsorption-growth-diffusion framework, this work establishes a route toward programmable defect architectures and large-area quantum materials engineered directly during synthesis.

\section{Methods}

\paragraph{Sample Fabrication.}
Monolayer V-doped WS$_2$ flakes were synthesized by a liquid-assisted chemical vapor deposition (CVD) approach, as described in detail in previous reports~\citeA{zhang2020monolayer}. Ammonium metatungstate hydrate ((NH$4$)$6$H$2$W${12}$O${40}$·xH$2$O) and sodium cholate hydrate (C${24}$H${39}$NaO$_5$·xH$_2$O) were dissolved in deionized water to prepare the W precursor solution. In parallel, vanadyl sulfate (VO[SO$_4$]) was dissolved in deionized water to form the V precursor solution. The two solutions were subsequently combined and spin-coated onto SiO$_2$/Si substrates. The coated substrates were loaded into a quartz tube furnace, with sulfur powder positioned upstream as the chalcogen source. Under a continuous flow of ultrahigh-purity Ar carrier gas, the system was heated to 825~°C and maintained for 15~min to promote growth. After the reaction, the furnace was allowed to cool naturally to room temperature under Ar atmosphere. The vanadium doping level was controlled by adjusting the volume ratio between the W and V precursor solutions.

\paragraph{Synchrotron X-ray Nanoprobe Microscopy.}
X-ray fluorescence microscopy was performed at the CARNAUBA beamline of the Sirius synchrotron light source (LNLS, Brazil)~\citeA{tolentino2023carnauba}. The samples were excited using a nanofocused X-ray beam (500 $\times$ 200~nm$^{2}$) under ambient conditions. An incident energy of 11~keV was used to excite the tungsten $L$-edges, while 6~keV provided optimal excitation of sulfur and vanadium $K$-edges. To avoid unnecessary beam exposure, a fast shutter system ensured the sample was exposed to the beam only during data acquisition.
Sample localization was facilitated by pre-marking the SiO$_2$ substrates with copper tape using the optical microscopy facilities at the Microscopic Samples Laboratory (LAM-LNLS). The fluorescence signal was collected using a silicon drift detector (SDD) in a 90$^{\circ}$ geometry relative to the incident beam. 

Regions of interest were initially identified using coarse fly-scans covering 250$\times$250~$\mu$m$^2$ areas with a 5~$\mu$m step size and 30~ms integration time. Once the target flakes were located, high-resolution hyperspectral maps were acquired in step-scan mode. For highly V-doped samples, we mapped areas of 250$\times$250~$\mu$m$^2$ with a 2.5~$\mu$m step size; for pristine samples, we used finer scans of 60$\times$60~$\mu$m$^2$ with a 0.6~$\mu$m step size. In both cases, the integration time was 0.5~s per pixel.
Element-specific maps were produced by integrating the characteristic emission lines: S-$K_\alpha$ (2.31 keV), V-$K_\alpha$ (4.95 keV), and W-$L_\alpha$ (8.40 keV), using the PyMCA software~\citeA{sole2007multiplatform}.

\paragraph{Quantitative Concentration Framework.}
All XRF maps were recorded under identical experimental conditions, including incident photon energy, flux, beam geometry, and integration time. Under these fixed conditions, the integrated fluorescence intensity of a given elemental emission line is proportional to the local area atomic concentration. For the monolayer geometry studied here, matrix and self-absorption effects are negligible, allowing direct ratiometric comparison of intensities across the sample. 

We assume that V incorporates exclusively through substitution on W sites. Accordingly, at any position $p$ on the metal sublattice, the local concentrations satisfy: $C^{W}(p) + C^{V}(p) = 100$. 

For any element $X$, and any two points $p_1$ and $p_2$, measured under identical experimental conditions, the corresponding concentrations scale with the measured XRF intensities,

\begin{equation}
    \frac{C^{X}(p_1)}{C^{X}(p_2)}=\frac{I_{X}(p_1)}{I_{X}(p_2)}.    
\end{equation}

Applying this to V and W yields

\begin{align}
    \frac{C^{V}(p_1)}{C^{V}(p_2)} &= \frac{I_{V}(p_1)}{I_{V}(p_2)}, \label{eqpropV} \\
    \frac{C^{W}(p_1)}{C^{W}(p_2)} &= \frac{I_{W}(p_1)}{I_{W}(p_2)}. \label{eqpropW}
\end{align}

Substituting $C^{W}=100-C^{V}$ into the second expression and eliminating $C^{V}(p_2)$ using the first yields the expression used in the main text,

\begin{equation}
    C^{V}(p_1) = 100 \times \frac{I_{V}(p_1) \left[I_{W}(p_1)-I_{W}(p_2) \right]}{I_{V}(p_2)I_{W}(p_1) - I_{V}(p_1)I_{W}(p_2)},
\end{equation}

Values of $C^{V}(p_1)$ were computed using several off-bisector positions $p_2$; the reported concentrations correspond to the mean and standard deviation across all reference points.

Sulfur occupies the chalcogen sublattice independently and was quantified via one-point normalization:

\begin{equation}
    C^{S}(p) = 100 \times \frac{I_{S}(p)}{I_{S,\mathrm{ref}}},
\end{equation}

\noindent where $I_{S,\mathrm{ref}}$ is the mean S-($K_{\alpha}$) intensity of pristine WS$_2$ monolayer. The relative S-vacancy fraction is estimated as $100\% - C^{S}(p)$.

The uncertainties for $C^{W}$, $C^{V}$, and $C^{S}$ combine counting statistics, spatial variability of the pristine reference, and the spread among different $p_2$ values.

\paragraph{Adsorption-Growth-Diffusion (AGD) Model.}
The spatial dopant distribution was modeled by solving the time-dependent diffusion equation for a triangular domain subject to preferential corner incorporation. As detailed in Supplementary Note~4, the concentration field follows Fick’s second law, where the local dopant concentration evolves through thermal diffusion with diffusion coefficient $D$.

\begin{equation}
    \frac{\partial}{\partial t}C({\mathbf r},t) = D \nabla^{2}C({\mathbf r},t),
\end{equation}

\noindent where $C({\mathbf r},t)$ is the local impurity. We consider a growth regime in which the crystal grows with a constant velocity $v$ until it reaches a final size $L$. Dopant incorporation occurs through two concurrent pathways. First, homogeneous adsorption along the crystal edges produces a uniform background concentration level, denoted $B$. Second, preferential incorporation at the three crystallographic vertices acts as localized sources with amplitude $A$.

Following the derivation in Supplementary Note~4, the concentration field generated by a single vertex source is expressed using a dimensionless diffusion function ${\mathcal I}_{\beta}$ :

\begin{equation}
    {\mathcal I}_{\beta} (X,Y) = \int_{0}^{1} \frac{1}{(1-Y^{\prime} + \eta)} \exp\left[{-\beta\frac{(X^2 +(Y-Y^{\prime})^2)}{(1-Y^{\prime} + \eta)}}\right] dY^{\prime},
\end{equation}

Coordinates $(X,Y)$ are normalized by the center-to-corner distance $L$, and $\eta$ is a dimensionless regularization parameter representing the finite source size. The single dimensionless parameter $\beta = {vL}/{4D}$ captures the competition between crystal growth velocity and thermal diffusion rates.

The total dopant concentration field is constructed as a linear superposition of three such contributions, corresponding to the three crystallographic vertices of the triangular domain. Explicitly, the concentration field is given by

\begin{equation} 
\begin{aligned} 
    C(X,Y) = & A\Bigg[ \mathcal{I}_{\beta}(X,Y) + \mathcal{I}_{\beta}\left( -\frac{X}{2} + \frac{\sqrt{3}Y}{2}, -\frac{\sqrt{3}X}{2} - \frac{Y}{2} \right) \\ & +\, \mathcal{I}_{\beta}\left( -\frac{X}{2} - \frac{\sqrt{3}Y}{2}, \frac{\sqrt{3}X}{2} - \frac{Y}{2} \right)\Bigg] + B.    
\end{aligned} 
\end{equation}

This expression naturally generates three symmetry-related high-density channels aligned along the crystallographic bisectors. When projected onto one-dimensional profiles, the solution reduces to the compact form used in the main text, $C(r) = A \mathcal{F}_\beta(r) + B$, where $\mathcal{F}_\beta(r)$ represents the effective diffusion profile governed by $\beta$.

Experimental concentration profiles extracted from XRF measurements were fitted to this expression to extract $A$, $B$, and $\beta$. The best agreement with experiment is obtained for $\beta \approx 50$, indicating a kinetically limited regime where the crystal growth velocity significantly exceeds the diffusive relaxation rate, resulting in the formation of dopant segregation channels.

\paragraph{Hyperspectral Raman Spectroscopy.}
Raman measurements were performed at room temperature using a WITec Alpha300 confocal microscope with 532~nm excitation laser and a sensitive charge-coupled device (CCD) at the LCPNano facilities at the Universidade Federal de Minas Gerais. The laser beam was focused onto the sample using a 100$\times$ Zeiss objective (NA $=$ 0.9), producing a diffraction-limited spot size of approx. $\sim$360~nm. To avoid laser-induced local heating or photo-degradation, the excitation power was limited to $\sim$200~$\mu$W at the sample surface. 

Hyperspectral images were acquired by raster-scanning the sample using a high-precision piezoelectric stage. For the high-resolution maps shown in Fig.~\ref{fig_Raman}c-f, we utilized a step size of 1~$\mu$m with an integration time of 0.5~s per pixel. The scattered light was collected in a backscattering geometry and dispersed by a 1800 grooves/mm grating, yielding a spectral resolution of better than 1~cm$^{-1}$~\citeA{sergio2023}.
Raw spectra were processed using the PortoFlow (FabNS) software. Background subtraction was applied to each spectrum, and phonon modes were fitted with Lorentzian line shapes to extract peak positions, linewidths, and intensities. The integrated intensity maps shown in Figs.~\ref{fig_Raman}c,e,g were normalized by the integrated intensity maps of the Si Raman peak (520~cm$^{-1}$) to correct possible variations in laser power, focus, and alignment.

To separate the strain and charge-doping contributions to the phonon renormalization, the frequencies of the $E^{\prime}$ and $A_1^{\prime}$ modes were projected onto the strain ($\varepsilon$) and doping ($n$) axes using the linear relations~\citeA{lee2023suppression}:

\begin{align}
    & \Delta n (10^{13}~\text{cm}^{-2}) = 0.51 \times \Delta \omega_{E^{\prime}} - 0.94 \times \Delta \omega_{A_1^{\prime}}, \\ 
    & \Delta \varepsilon (\%) = - 0.27 \times \Delta \omega_{E^{\prime}} + 0.04 \times \Delta \omega_{A_1^{\prime}}.
\end{align}

\noindent Here $\Delta \omega$ represents the frequency shift relative to the pristine monolayer (in cm$^{-1}$). Applying this transformation pixel-by-pixel to the hyperspectral dataset yielded the spatial strain map shown in Fig.~\ref{fig_Raman}h. For the characteristic redshift observed at the bisector lines ($\Delta \omega_{E^{\prime}} \approx -3~\text{cm}^{-1}$ and $\Delta \omega_{A_1^{\prime}} \approx -2.7~\text{cm}^{-1}$), this calculation yields a local tensile strain of $\Delta \varepsilon \approx 0.70\%$ along the segregation channels.

\paragraph{DFT Calculations.}

First-principles calculations were performed within density functional theory (DFT) using the Vienna Ab initio Simulation Package (VASP). The projector-augmented wave (PAW) method~\citeA{PhysRevB.50.17953,PhysRevB.59.1758,PhysRevB.54.11169,KRESSE199615} was employed, and exchange–correlation effects were described within the generalized gradient approximation using the Perdew–Burke–Ernzerhof (PBE) functional~\citeA{perdew1996generalized}. Electronic correlations in the transition-metal states were treated within the DFT$+U$ formalism, with an effective Hubbard parameter $U_{\mathrm{eff}} = 3.0$~eV applied to the vanadium $d$ orbitals~\citeA{Yun2020,Duong2019}. A plane-wave basis set with an energy cutoff of 600~eV was used, and total energies were converged to $10^{-6}$~eV. A vacuum spacing of at least 13~\AA\ was introduced along the out-of-plane direction to avoid spurious interactions. Spin–orbit coupling (SOC) was included self-consistently in all final property calculations.

Ground-state atomic configurations and substitutional disorder were determined using the USPEX evolutionary algorithm~\citeA{uspex1,uspex2,uspex3}. For each stoichiometry in the $3 \times 3$ supercell ($x \approx 0.11, 0.22, 0.44$), we carried out an extensive configurational search in which a permutation operator was applied concurrently to both the atomic occupations (V and W sites) and the initial magnetic configurations, to minimize the total enthalpy. This combined configurational–magnetic permutation scheme enabled efficient sampling of the coupled structural–magnetic energy landscape and facilitated the identification of the specific cation ordering and spin arrangement that yield the global enthalpy minimum. During the evolutionary search, spin-polarized calculations were conducted without inclusion of spin–orbit coupling (SOC) to determine the magnetic ground state in a computationally efficient manner. Each generation comprised 30 candidate structures, and convergence was defined as the persistence of the lowest-enthalpy individual for at least 20 successive generations. In all cases, the lowest-enthalpy configurations converged to a ferromagnetic low-spin (FM-LS) ground state.

To evaluate the local strain fields and the interface between V-rich and pristine regions, we independently constructed a periodic stripe-patterned WS$_2$ supercell with dimensions $4 \times 12$. In this model, V atoms were randomly distributed within a central V-rich WS$_2$ stripe, which is periodically connected to adjacent pristine WS$_2$ domains. This geometry provides a simplified representation of the experimentally observed dopant-rich bisectors while preserving full periodic boundary conditions. It also enables a spatially resolved analysis of the effective metal-sulfur bond-length distribution and of the propagation of lattice disorder across the V-rich/pristine WS$_2$ interface. All structures, including the USPEX-derived configurations and the periodic
supercell containing the disordered V-rich WS$_2$ stripe, underwent final electronic refinement in VASP using SOC and a dense $\mathbf{\Gamma}$-centered $\mathbf{k}$-point mesh.

%The relative stability of the doped systems was evaluated through the formation energy ($E_f$), defined as:$$ E_f = E_{doped} - \sum n_i \mu_i$$
%where $E_{doped}$ represents the total energy of the $3 \times 3 \times 1$ supercell for each respective phase (2H or 1T), $n_i$ is the number of atoms of species $i$, and $\mu_i$ is its corresponding chemical potential. To ensure a realistic representation of the synthesis conditions, the chemical potentials were derived from the total energies of the respective bulk ground states: BCC Tungsten (W), BCC Vanadium (V), and molecular sulfur in its gaseous diatomic form (S$_2$). The resulting energy difference confirms the 2H phase as the stable host for the observed deterministic dopant and strain architectures.

Phonon modes at the $\mathbf{\Gamma}$-point were obtained within density functional perturbation theory (DFPT). The macroscopic dielectric tensor was computed within the same linear-response framework. Non-resonant Raman activities were evaluated from the derivatives of the macroscopic dielectric tensor with respect to the normal mode coordinates, $\partial \varepsilon / \partial Q$, following the formalism of Porezag and Pederson~\citeA{Porezag1996}. Raman intensities were evaluated within the Placzek approximation in the non-resonant limit using the \texttt{vasp\_raman.py} implementation~\citeA{raman_sc_vasp}. For these final properties, Brillouin zone integrations were carried out using a $\mathbf{\Gamma}$-centered $10\times10\times1$ mesh.
Phonon eigenvectors were visualized using the Phonon Website developed by Miranda~\citeA{phononwebsite}.

\backmatter

\section*{Supplementary information}

Supplementary Notes~1--9 and Figs.~S1--S14.

\section*{Acknowledgements}

F.B.S., G.E.M., and M.D.T. acknowledge support from FAPESP (2024/15969-1).
F.B.S., R.O., M.J.S.M., B.R.C., I.F.C., F.M., G.E.M., L.M.M., H.C., and M.D.T. acknowledge financial support from the Brazilian agencies CNPq and CAPES.
M.J.S.M., B.R.C., L.M.M., and H.C. thank FAPEMIG, FINEP, the National Institute of Science and Technology (INCT) in Carbon Nanomaterials, and the Rede Mineira de Materiais 2D (FAPEMIG) for funding.
M.J.S.M. and I.F.C. also acknowledge support from the Universidade Federal de Ouro Preto. The authors also acknowledge the support received from the National Laboratory for Scientific Computing (LNCC/MCTI, Brazil) at São Paulo (CENAPAD-SP).
B.R.C. acknowledges additional support from the Universidade Federal do Rio Grande do Norte (UFRN).
Z.Y., M.L., E.G.H., and M.T. acknowledge financial support from AFOSR (FA9550-23-1-0447).

This research used the facilities of the Microscopic Samples Laboratory (LAM) and the CARNAUBA beamline at Sirius, part of the Brazilian Synchrotron Light Laboratory (LNLS), under proposal No. 20250170. The LNLS is integrated into the Brazilian Center for Research in Energy and Materials (CNPEM), a private non-profit organization under the supervision of the Brazilian Ministry for Science, Technology, and Innovation (MCTI). The authors also thank LCPNano for facility access and FabNS for access to the PortoFlow software used for data analysis.

\section*{Contributions}
F.B.S., R.O., and L.M.M conceived the ideas. 
F.B.S. and R.O. performed XRF microscopy experiments. 
F.B.S. and F.M. conducted Raman spectroscopy experiments. 
E.G.H., Z.Y., M.L., and M.T. synthesized the samples and performed AFM experiments. 
F.B.S. analyzed the experimental data. 
H.C. developed the AGD model. 
M.J.S.M. and I.F.C. performed DFT calculations. 
F.B.S., R.O., M.J.S.M., B.R.C., and H.C. interpreted the results and wrote the manuscript with inputs from all authors. 
F.B.S. and M.D.T. led the project. 
All authors discussed the results and revised the manuscript.

\section*{Declarations}

The authors declare no competing interests.


%\bibliography{References}% common bib file


\clearpage

\setcounter{figure}{0}
\renewcommand{\figurename}{\text{Supplementary Figure}}
\renewcommand{\thefigure}{\text{S\arabic{figure}}}

\renewcommand{\thetable}{\text{S\arabic{table}}}

\setcounter{equation}{0}
\renewcommand{\theequation}{S\arabic{equation}}

%\setstretch{1.2}

% -------------------- Title --------------------

\title{\textbf{\Large{Supplementary Information for}} \\
\\
{\Large Anisotropic Dopant and Strain Architectures in WS$_2$ Nanocrystals Driven by Growth Kinetics}
\\
\\
{\noindent
\large Frederico B. Sousa, Raphaela de Oliveira, Matheus J. S. Matos, Elizabeth Grace Houser, Igor Ferreira Curvelo, Zhuohang Yu, Mingzu Liu, Felipe Menescal, Gilmar Eugenio Marques, Leandro M. Malard, Mauricio Terrones, Bruno R. Carvalho, Helio Chacham, and Marcio D. Teodoro
}}

\date{}

\maketitle

\vspace{5mm}

\noindent
This PDF file includes:

\begin{itemize}
  \item Supplementary Note 1: X-ray fluorescence spectroscopy analysis
  \item Supplementary Note 2: Quantitative estimation of vanadium concentration
  \item Supplementary Note 3: Additional X-ray fluorescence data and reproducibility
  \item Supplementary Note 4: The Adsorption-Growth-Diffusion (AGD) Model
  \item Supplementary Note 5: Phase analysis
  \item Supplementary Note 6: Additional Raman spectroscopy data
  \item Supplementary Note 7: Topographic analysis
  \item Supplementary Note 8: Vibrational calculations
  \item Supplementary Note 9: Doping engineering via temporal flux modulation
\end{itemize}

\newpage

% -------------------- Supplementary Notes --------------------

\section*{\large{Supplementary Note 1: X-ray fluorescence spectroscopy analysis}}

\begin{figure}[!htbp]
	\centering
	\includegraphics[width=1.0\linewidth]{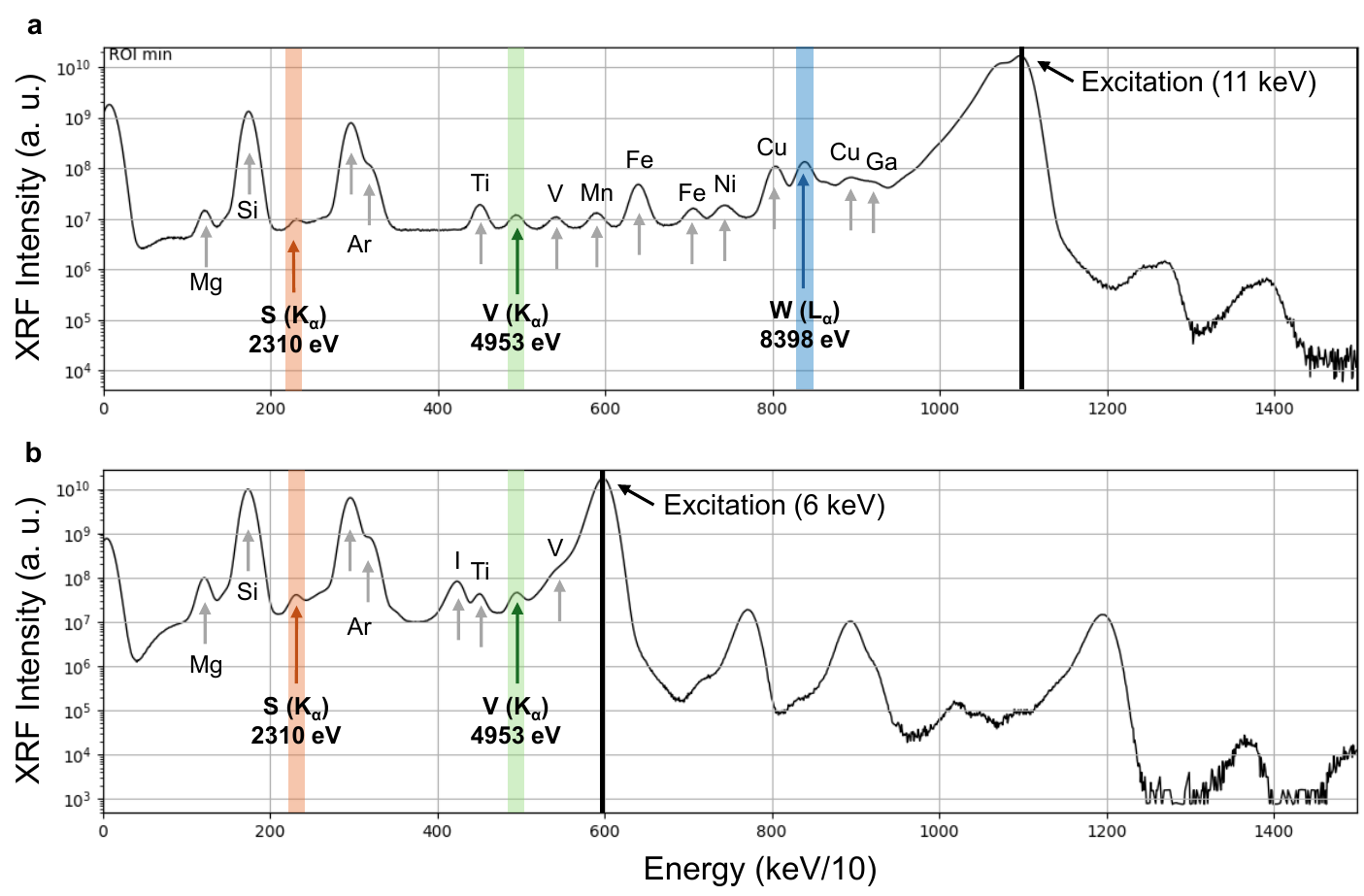} 
	\caption{ {\bf Synchrotron X-ray fluorescence spectra.}
    Representative XRF spectra for a V-doped WS$_2$ monolayer obtained at (\textbf{a}) 11~keV (\textbf{b}) and 6~keV  incident photon energies. For each resolved peak, the corresponding atomic species and transition are indicated. For S, V, and W emissions analyzed in this study, shaded vertical strips represent the 200~eV integration windows used to extract the local fluorescence intensities. The ratiometric comparison of these integrated areas allows for the quantitative determination of the local atomic concentrations presented in Figure~1 of the main text.}
	\label{SuppFig1}
\end{figure}

Supplementary Figure~\ref{SuppFig1} shows representative X-ray fluorescence (XRF) spectra acquired from a highly vanadium-doped WS$_2$ monolayer using incident beam energies of 11~keV and 6~keV.

Several characteristic peaks corresponding to distinct atomic species are resolved. In particular, the S-$K_\alpha$ (2310~eV), V-$K_\alpha$ (4953~eV), and W-$L_\alpha$ (8398~eV) emissions--which form the basis of our chemical mapping--are highlighted by orange, green, and blue vertical stripes, respectively. To quantify the elemental distribution, we integrated the spectral area within a 200~eV window centered on each emission peak (denoted by the stripe widths). This integration procedure was performed pixel-by-pixel across the raster-scanned area to generate the high-resolution intensity maps shown in Figure~1 of the main text.

While other transitions are present within this energy range, such as W-$L_\beta$ (9672~eV) and V-$K_\beta$ (5428~eV), they were excluded from the quantitative mapping due to partial convolution with the primary excitation peaks or adjacent environmental signals. The additional features observed in the spectra are attributed to the experimental environment: silicon (Si) and copper (Cu) signals originate from the substrate and mounting tape, argon (Ar) arises from the ambient atmosphere, and metallic traces (e.g., Ti, Mn, Fe, Ni, and Ga) are associated with the internal components of the XRF setup or the beamline hutch.

%%%%%%%%%%%%%%%%%%%%%%
%\newpage
\section*{\large{Supplementary Note 2: Quantitative estimation of vanadium concentration}}
% precisa explicar porque nao precisamos de uma amostra padrao: W como referencia interna assumindo que temos apenas substituicao de dopagem.

The quantitative determination of absolute atomic concentrations from X-ray fluorescence (XRF) intensities typically requires external standards or complex matrix corrections. However, for the monolayer geometry studied here, where self-absorption is negligible, we developed a self-consistent ratiometric framework that utilizes the tungsten host lattice as an internal reference.

We assume vanadium incorporates into the WS$_2$ lattice exclusively by substituting for tungsten atoms ($V_W$). Consequently, at any point $p$ on the metal sublattice, the local atomic concentrations must satisfy:

\begin{equation}
    C^W(p)+C^V(p) = 100\%
\end{equation}

Under identical experimental conditions, the measured fluorescence intensity $I_X(p)$ of an element $X$ is directly proportional to its local concentration $C^X(p)$. By comparing the intensities at two distinct points, $p_1$ and $p_2$, we obtain the following relations:

\begin{equation}
    \frac{C^V(p_1)}{C^V(p_2)} = \frac{I_V(p_1)}{I_V(p_2)} \quad \text{and} \quad \frac{C^W(p_1)}{C^W(p_2)} = \frac{I_W(p_1)}{I_W(p_2)}
\end{equation}

By substituting the constraint $C^W = 100 - C^V$ into these relations, we derive the expression for the absolute vanadium concentration at point $p_1$ (Equation~2 of the main text):

\begin{equation}
    C^{V}(p_1) = 100 \times \frac{I_{V}(p_1) \left[I_{W}(p_1)-I_{W}(p_2) \right]}{I_{V}(p_2)I_{W}(p_1) - I_{V}(p_1)I_{W}(p_2)},
%    \label{eqCV}
\end{equation}

To improve the statistical robustness of the 33\% peak concentration reported in the manuscript, we utilized multiple reference points ($p_2$) to independently estimate the concentration at a fixed bisector position ($p_1$, white circle in Supplementary Fig.~\ref{SuppFig2}). As shown in Supplementary Fig.~\ref{SuppFig2}a-b, $p_1$ was fixed at the peak of the segregation channel, while $p_2$ was varied along a transverse line (red dashed arrow).  

The resulting distribution of $C^V(p_1)$ values for different $p_2$ reference positions is summarized in Supplementary Figure~\ref{SuppFig2}c. This analysis yields a stable mean vanadium concentration of (33$\pm$3)\% at the bisector peak. Once this absolute reference value is established, the concentration at any other pixel in the map is calculated by a direct comparison of V-$K_{\alpha}$ intensities.

\begin{figure}[!htbp]
	\centering
	\includegraphics[width=1.0\linewidth]{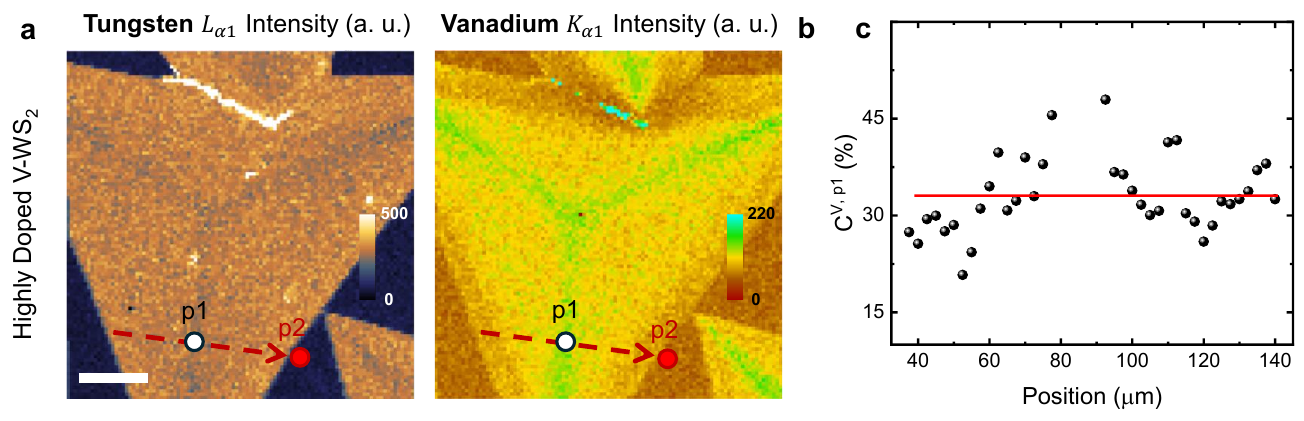} 
	\caption{{\bf Statistical validation of the ratiometric concentration framework.}
    \textbf{a,b,} XRF intensity maps for the W-$L_{\alpha}$ and V-$K_{\alpha}$ transitions in a highly doped domain. The white circle denotes the fixed point $p_1$ (on-bisector), while the red circle represents the variable reference point $p_2$, which is scanned along the red dashed arrow to ensure statistical convergence. Scale bar: 50~$\mu$m.
    \textbf{c,} Calculated vanadium concentration at $p_1$ ($C^V(p_1)$) as a function of the $p_2$ reference position. Black dots represent individual ratiometric estimates; the red horizontal line indicates the mean value of (33$\pm$3)\%, confirming the reliability of the substitutional model.}
	\label{SuppFig2}
\end{figure}

%%%%%%%%%%%%%%%%%%%%%%

%\newpage
\section*{\large{Supplementary Note 3: Additional X-ray fluorescence data and reproducibility}}

To evaluate the reproducibility of the observed segregation and assess the sensitivity of our chemical mapping, we performed additional XRF microscopy experiments on pristine, moderately doped, and highly V-doped WS$_2$ monolayers.

\begin{figure}[!htbp]
	\centering
	\includegraphics[width=1.0\linewidth]{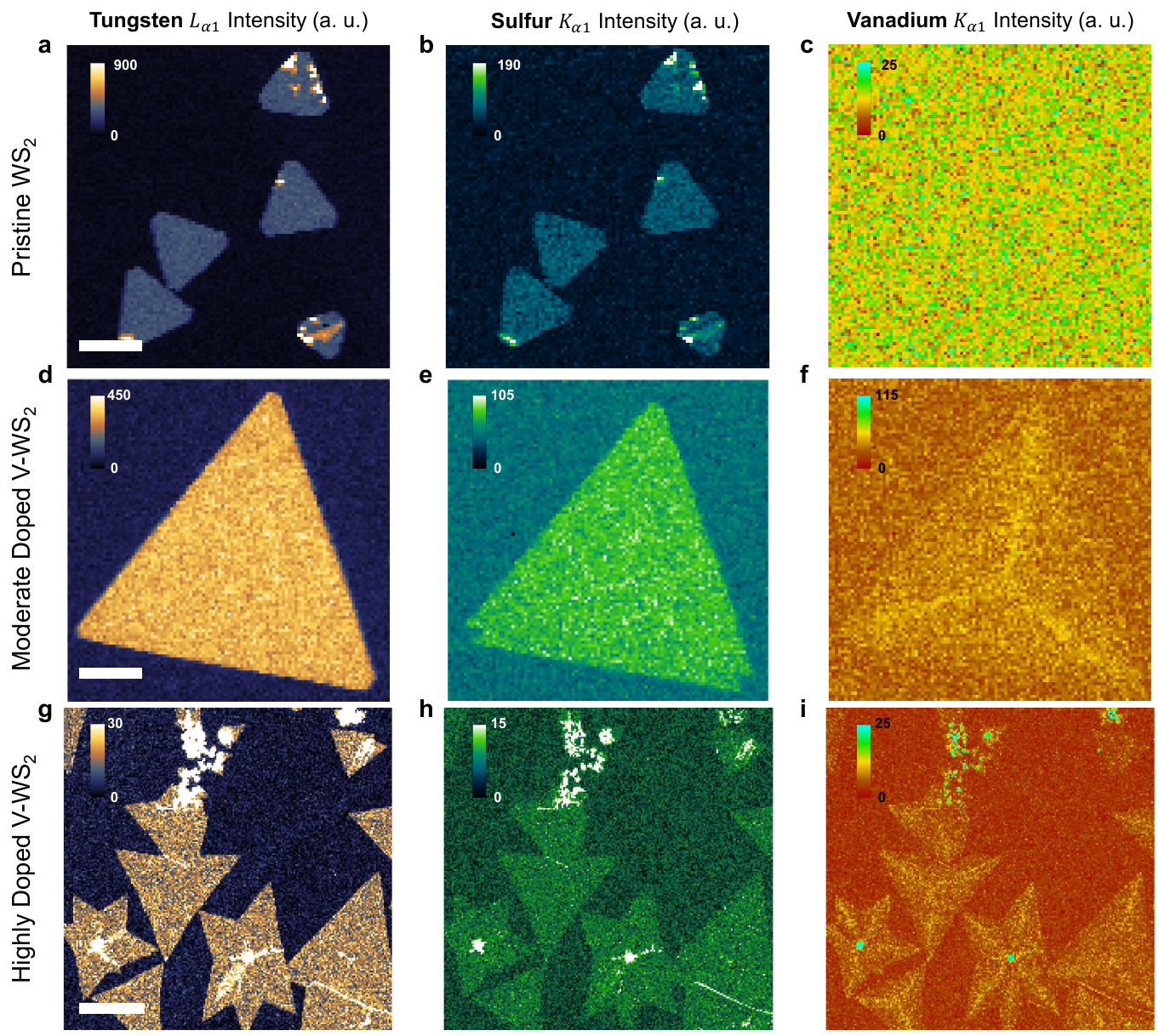} 
	\caption{{\bf XRF mapping across different doping regimes.}
    XRF intensity maps for the W-$L_{\alpha}$ (\textbf{a,d,g}), S-$K_{\alpha}$ (\textbf{b,e,h}), and V-$K_{\alpha}$ (\textbf{c,f,i}) transitions. 
    \textbf{a-c,} Pristine WS$_2$ showing a uniform host lattice and negligible vanadium signal.
    \textbf{d-f,} Moderately doped monolayer highlighting the sensitivity of the technique to dilute vanadium populations ($\sim$3-7\%).
    \textbf{g-i,} Survey maps of multiple highly doped domains connecting the domain centers to the vertices, confirming the reproducibility of the segregation mechanism. Scale bars: 50~$\mu$m (\textbf{a}), 15~$\mu$m (\textbf{d}), and 200~$\mu$m (\textbf{g}).}
	\label{SuppFig3}
\end{figure}

Supplementary Figures~\ref{SuppFig3}a-c show the XRF intensity maps for a pristine domain. As expected, the tungsten and sulfur distributions are spatially uniform, while the vanadium signal remains at the background level. In the case of the moderately doped WS$_2$ monolayer (Supplementary Figs.~\ref{SuppFig3}d-f), the host lattice (W and S) remains largely homogeneous. However, the V-$K_{\alpha}$ map reveals a clear enrichment along the bisector lines. Applying our ratiometric framework, we estimated vanadium concentrations of approximately 7\% at the bisector and 3\% at the domain edges. These results demonstrate that synchrotron-based XRF mapping is sufficiently sensitive to resolve dopant heterogeneities down to a few atomic percent, even in the limit of a single atomic layer. For this moderate doping regime, the corresponding tungsten depletion is not resolved because the reduction in W-$L_{\alpha}$ intensity falls within the statistical noise of the host lattice signal. 

Finally, Supplementary Fig.~\ref{SuppFig3}i displays vanadium maps for several additional domains, all of which exhibit characteristic enrichment along the crystallographic bisectors. This confirms that anisotropic dopant segregation is a fundamental and reproducible feature of the growth process. While the tungsten and sulfur signals in these wide-field maps remain uniform (Supplementary Figs.~\ref{SuppFig3}g,h), the localized W-depletion at the bisectors--observed in the high-resolution maps of Figure~1--is not resolved here. This is due to the lower spatial resolution and larger step size employed to survey multiple fakes across a wider area of the substrate.

%%%%%%%%%%%%%%%%%%%%%%
%\newpage
\section*{\large{Supplementary Note 4: The Adsorption-Growth-Diffusion (AGD) Model}}

As discussed in the manuscript, we expect that, as a triangular flake grows by the adsorption of atoms, the corner regions will contain, at a given instant of growth, a larger concentration of defects.  As growth proceeds, three lines with a high concentration of defects appear on the growing flake, tracing the trajectories from the center to the corners as illustrated in Supplementary Fig.~\ref{SuppFig4}.

\begin{figure}[!htbp]
	\centering
	\includegraphics[width=0.4\linewidth]{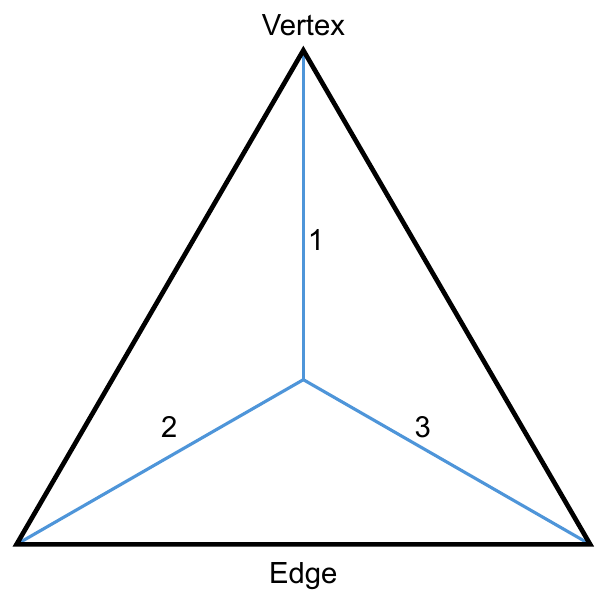} 
	\caption{{\bf Geometric model of deterministic dopant incorporation.}
    Schematic representation of a triangular WS$_2$ flake during the growth process. The black outline denotes the expanding domain boundaries. The blue lines represent the trajectories of high dopant concentration originating from the preferential incorporation of impurities at the three domain vertices. These trajectories propagate from the nucleation center ($x=0$, $y=0$) toward the vertices as the flake grows at a constant velocity $v$.}
	\label{SuppFig4}
\end{figure}

To extend this reasoning to a theory capable of quantitatively predicting the experimental findings, we consider that at a given moment of growth, the triangular flake has $N_{c}$ corner sites (which remain constant during growth) and $N_{e}$ lateral edge sites (which increase as the flake expands). We define $E_{f}^{c}$ and $E_{f}^{e}$ as the impurity formation energies at corner and edge sites, respectively. Under conditions of thermodynamic equilibrium, the probability $P_{corner}$ that an impurity is adsorbed at a corner site is given by:

\begin{equation}
   P_{corner} = \frac{N_{c}e^{- E_{f}^{c}/kT}}{N_{c}e^{- E_{f}^{c}/kT} + N_{e}e^{- E_{f}^{e}/kT}} = \frac{1}{1 + \frac{N_{e}}{N_{c}}e^{- \left( E_{f}^{e} - E_{f}^{c} \right)/kT}}.
\end{equation}

Defining $y$ as the length of the bisector line (e.g., line labeled as $1$ in Supplementary Fig.~\ref{SuppFig4}), and assuming $y = 0$ corresponds to a seed for which there are only corner sites, we obtain $N_{e} = y/a$. Therefore, we find:

\begin{equation}
    P_{corner} = \frac{1}{1 + \alpha y}, \quad \text{where} \quad \alpha = \frac{e^{- \left( E_{f}^{e} - E_{f}^{c} \right)/kT}}{aN_{c}}.
\end{equation}

The parameter $\alpha$ is constant during growth if the temperature is kept constant. We now consider the effect of impurity diffusion away from these defect lines. At this stage of the derivation, we consider the contribution of corner-selective doping, which is modeled as a continuous sequence of infinitesimal $\delta$-doping events along the center-to-corner trajectories. These incorporation events progress with the characteristic growth velocity $v$ of the domain vertices. 
We assume that the diffusion is governed by the second Fick's law:

\begin{equation}
    \frac{\partial}{\partial t}C\left( \mathbf{r},t \right) = D\mathbf{\nabla}^{2}C\left( \mathbf{r},t \right),
\end{equation}

\noindent where $D$ is the thermally activated diffusion constant and $C\left(\mathbf{r},t \right)$ is the impurity concentration at a given position and time. For a $\delta$-doping event occurring at position $\mathbf{r}_{0}$ (at a distance $y^\prime$ from the flake center) and time $t_{0}$, the initial concentration is weighted by the adsorption probability:

\begin{equation}
    C\left( \mathbf{r}{,\ t}_{0} \right) = \left( \frac{C_{0}}{1 + \alpha y^{\prime}} \right)\delta\left( \mathbf{r} - {\mathbf{r}}_{0} \right),
\end{equation}

\noindent where $C_{0}$ is a constant. In a two-dimensional material, the time evolution for at $t > t_{0}$, results in:

\begin{equation}
    C\left( \mathbf{r},t \right) = \left( \frac{C_{0}}{1 + \alpha y^{\prime}} \right)\frac{1}{4\pi D\left( t - t_{0} \right)}e^{- \left| \mathbf{r} - {\mathbf{r}}_{0} \right|^{2}/4D\left( t - t_{0} \right)}.
\end{equation}

From the time evolution of the concentration profile, we can define a characteristic diffusion length $l = \sqrt{D\left( t - t_{0} \right)}$, which represents the spatial extent of dopant redistribution away from the initial incorporation site. We assume that the growth process initiates at time $t = 0$ and stops at $t = t_{g}$, at which point the precursor sources are terminated, and the temperature is rapidly quenched, effectively halting the impurity diffusion process. Throughout the synthesis, the triangular domain is assumed to expand at a constant velocity $v$, defined as the radial velocity of the domain vertices relative to the flake center.

We first consider the impurity diffusion process originating from the defect trajectory defined as line 1 in Supplementary Fig.~\ref{SuppFig4}. Due to the $C_3$ symmetry of the triangular domain, the diffusion processes originating from the lines $2$ and $3$ are equivalent through rotations of $\pm 2\pi/3$ and will be superposed to the total concentration profile later on. 

Line $1$ is defined as growing from the domain center ($x = 0$, $y = 0$) toward the final upper vertex at ($x = 0$, $y = vt_{g}$). At the conclusion of the synthesis ($t_g$), a doping event that occurred at the a position $y^{\prime}$ along this trajectory has undergone diffusion for a duration of $t = t_{g} - y^{\prime}/v$. Consequently, the differential contribution of this specific incorporation event to the mesoscopic impurity distribution is given by:

\begin{equation}
    dc = \left( \frac{C_{0}}{1 + \alpha y^{\prime}} \right)\frac{1}{4\pi Dvt_{g}\left( t_{g} - y^{\prime}/v \right)}e^{- \left\lbrack x^{2} + \left( y - y^{\prime} \right)^{2} \right\rbrack/\left\lbrack 4D\left( t_{g} - y^{\prime}/v \right) \right\rbrack}dy^{\prime}.
\end{equation}

And the total diffusion profile of line 1 at the end of the growth is

\begin{equation}
    C_{1}(x,y) = \frac{C_{0}}{4\pi Dvt_{g}}\int_{0}^{vt_{g}}{\frac{1}{(1 + \alpha y^{\prime})\left( t_{g} - y^{\prime}/v \right)}e^{- \left\lbrack x^{2} + \left( y - y^{\prime} \right)^{2} \right\rbrack/\left\lbrack 4D\left( t_{g} - y^{\prime}/v \right) \right\rbrack}dy^{\prime}}.
\end{equation}

Introducing dimensionless coordinates $X = \frac{x}{vt_{g}}$, $Y = \frac{y}{vt_{g}}$, $Y^{\prime} = \frac{y^{\prime}}{vt_{g}}$, along with the dimensionless kinetic parameters $\beta = \frac{v^{2}t_{g}}{4D}$ and $\gamma = vt_{g}\alpha = \frac{vt_{g}e^{- \left( E_{f}^{e} - E_{f}^{c} \right)/kT}}{aN_{c}}$.

By substituting these into the integration, the profile becomes:

\begin{equation}
    C_{1}(X,Y) = \frac{C_{0}}{4\pi Dt_{g}}\int_{0}^{1}{\frac{1}{\left( 1 + \gamma Y^{'} \right)\left( 1 - Y^{\prime} \right)}e^{- \beta\left\lbrack X^{2} + \left( Y - Y^{\prime} \right)^{2} \right\rbrack/\left\lbrack 1 - Y^{\prime} \right\rbrack}dY^{\prime}},
\end{equation}

\noindent where the definite integral, although non-expressible in terms of elementary analytical functions, can be calculated numerically. This result shows that the impurity concentration profile $C_{1}(X,Y)$ is fundamentally parametrized by the two dimensionless parameters $\beta$ and $\gamma$, alongside a normalization prefactor.

We now estimate the orders of magnitude for these parameters based on the experimental data:

\begin{itemize}
    \item \textbf{Order of magnitude of $\beta$}: For the studied V-doped WS$_2$ monolayers, the growth temperature is $T =$ 1098~K, the growth time is $t_{g} = 900$~s, and the center-to-vertex distance $vt_{g} \cong 1.2 \times 10^{- 4}$~m. This results in a growth velocity $v \cong 1.3 \times 10^{- 7}$~m/s. If we consider a diffusion length $l = \sqrt{Dt_{g}}$ of approximately $12\ \mu$m (roughly $1/10$ of $vt_{g}$), we estimate $10 < \beta < 100$.
    
    %In the experiments of Rosa et al.~\citeA{rosa2022investigation}, the growth temperature is $T =$ 1073~K, the growth time is $t_{g} = 630$~s, and the center-to-vertex distance $vt_{g} \cong 4 \times 10^{- 5}$~m. This results in a growth velocity $v \cong 6.3 \times 10^{- 8}$~m/s. If we consider a diffusion length $l = \sqrt{Dt_{g}}$ of approximately $4\ \mu$m (roughly $1/10$ of $vt_{g}$), we estimate $10 < \beta < 100$.

    \item \textbf{Order of magnitude of $\gamma$}: Considering a formation energy difference $E_{f}^{e} - E_{f}^{c} \cong 3.9$~eV (calculated for tungsten vacancies~\citeB{rosa2022investigation}), the resulting $\gamma\sim 10^{- 15}$ is negligible. Therefore, $\gamma$ can be set to zero for the purpose of fitting the experimental data.
\end{itemize}

The above discussion leads to a simplified version of the model. However, to account for the finite size of the vertex incorporation zone and to avoid the mathematical divergence (singularity) of the integral at the domain center ($Y^{\prime} \to 1$ and $X,Y \to 0$), we introduce a dimensionless regularization parameter $\eta$. This parameter physically represents the spatial extent of the corner site where dopants are initially adsorbed. The expression for the profile of line $1$ reads:

\begin{equation}
    C_{1}(X,Y) = A\ {\mathcal I}_{\beta}(X,Y) = A\int_{0}^{1}{\frac{1}{\left( 1 - Y^{\prime} + \eta \right)}e^{- \beta\left\lbrack X^{2} + \left( Y - Y^{\prime} \right)^{2} \right\rbrack/\left\lbrack 1 - Y^{\prime} + \eta \right\rbrack}dY^{\prime}},
\end{equation}

\noindent where the prefactor $A = \frac{C_{0}}{4\pi Dt_{g}}$ is a parameter to be fitted, representing the non-normalized count of vanadium atoms. In our analysis, we typically set $\eta \sim 10^{-2}$, which is consistent with the experimental spatial resolution.

To obtain the total concentration profile, we must include the contributions from lines $2$ and $3$, which are equivalent to line $1$ through rotations of $\pm 2\pi/3$. Additionally, we account for the background impurity adsorption at the flake edges, which results in a homogeneous, maximum-entropy doping profile represented by a fitting constant $B$. The final total concentration profile is given by:

\begin{equation} 
\begin{aligned} 
    C(X,Y) = & A\Bigg[ \mathcal{I}_{\beta}(X,Y) + \mathcal{I}_{\beta}\left( -\frac{X}{2} + \frac{\sqrt{3}Y}{2}, -\frac{\sqrt{3}X}{2} - \frac{Y}{2} \right) \\ & +\, \mathcal{I}_{\beta}\left( -\frac{X}{2} - \frac{\sqrt{3}Y}{2}, \frac{\sqrt{3}X}{2} - \frac{Y}{2} \right)\Bigg] + B.    
\end{aligned} 
\end{equation}

\noindent where $X = \frac{x}{L}$, $Y = \frac{y}{L}$, and $L$ is the center-to-vertex length, and

\begin{equation}
    {\mathcal I}_{\beta}(X,Y) = \int_{0}^{1}{\frac{1}{\left( 1 - Y^{\prime} +\eta \right)}e^{- \beta\left\lbrack X^{2} + \left( Y - Y^{\prime} \right)^{2} \right\rbrack/\left\lbrack 1 - Y^{\prime}  + \eta \right\rbrack}dY^{\prime}}.
\end{equation}

We can write this expression as $C(r) = A \mathcal{F}_\beta(r) + B$, where $\mathcal{F}_\beta(r)$ represents the effective diffusion profile governed by $\beta$, as presented in the main text.

\begin{figure}[!htbp]
	\centering
	\includegraphics[width=1.0\linewidth]{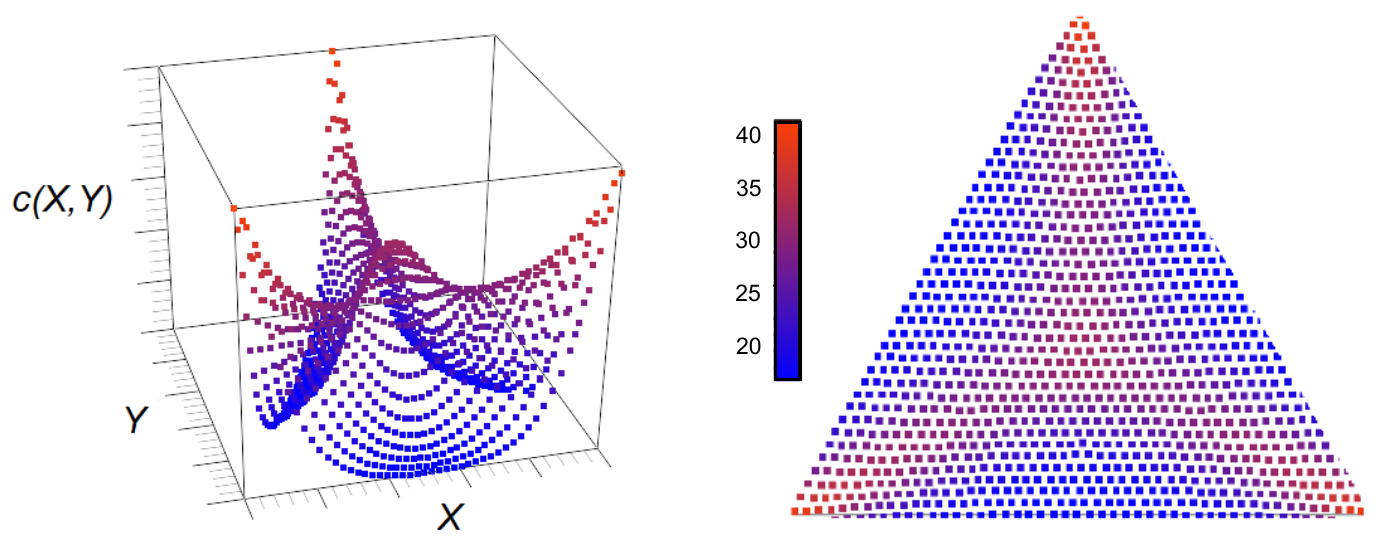} 
	\caption{{\bf Simulated dopant concentration profiles.}
    Numerical simulation of the total concentration profile $C(X,Y)$ derived from the AGD model using the dimensionless kinetic parameter $\beta = 25$. Left: 3D perspective view highlighting the ``hotspot'' at the domain center ($X = 0$, $Y = 0$) and the elevation at the domain edges. Right: Top-view 2D map illustrating the non-monotonic concentration along the bisector trajectories. The central maximum is a result of the constructive superposition of the three symmetry-equivalent diffusion fields, while the edge intensity is attributed to the shorter diffusion times available for dopants incorporated toward the end of the growth process ($t \to t_g$).}
	\label{SuppFig5}
\end{figure}

Supplementary Figure~\ref{SuppFig5} shows $C(X,Y)$ for $\beta = 25$ and $B = 0$. The simulation predicts a dopant concentration maximum at the domain center ($X = 0$, $Y = 0$) and at the triangle edges. Along the bisector trajectories, the behavior is non-monotonic, featuring a local minimum between the central and edge maxima. The central maximum arises from the constructive superposition of the three doping trajectories, resulting in a threefold intensity enhancement relative to a single line at the nucleation center.

%%%%%%%%%%%%%%%%%%%%%%

\newpage
\section*{\large{Supplementary Note 5: Phase analysis}}

To assess whether V incorporation destabilizes the host $2H$ lattice in favor of an octahedral coordination, we compared the total energies of pristine WS$_2$, the single-V substitutional configuration, and a two-V substitutional configuration in both the $2H$ and $1T$ phases. For pristine WS$_2$, the $2H$ phase is favored over the $1T$ phase by $0.28$ eV/atom, corresponding to $0.83$ eV per WS$_2$ formula unit~\citeB{han2021one}. Upon V incorporation, this energy separation decreases substantially, to $0.17$ eV/atom for the 1V configuration and $0.11$ eV/atom for the 2V configuration. Thus, V substitution partially stabilizes the octahedral coordination, but does not invert the thermodynamic ordering of the two polymorphs. This conclusion is consistent with the known phase hierarchy of group-VI transition-metal dichalcogenides, for which metastable $1T/1T'$ polymorphs generally require higher formation energies than the thermodynamically stable $2H$ phase and typically require external stabilization through charge transfer, electric fields, strain, epitaxial templating, or other non-equilibrium mechanisms~\citeB{Sokolikova2020,han2021one}. Although metastable $1T'$-WS$_2$ crystals and V-assisted $1T$-WS$_2$ monolayers have been demonstrated under dedicated synthetic conditions~\citeB{Lai2021,han2021one}, such phases are not the default thermodynamic products of WS$_2$ growth. Therefore, the Raman anomalies observed here are more consistently assigned to local symmetry lowering, defect-induced lattice frustration, and strain-mediated distortions within a $2H$-derived framework, rather than to a global $2H \rightarrow 1T/1T'$ transformation.

%%%%%%%%%%%%%%%%%%%%%%
%\newpage
\section*{\large{Supplementary Note 6: Additional Raman spectroscopy data}}

\begin{figure}[!htbp]
	\centering
	\includegraphics[width=0.85\linewidth]{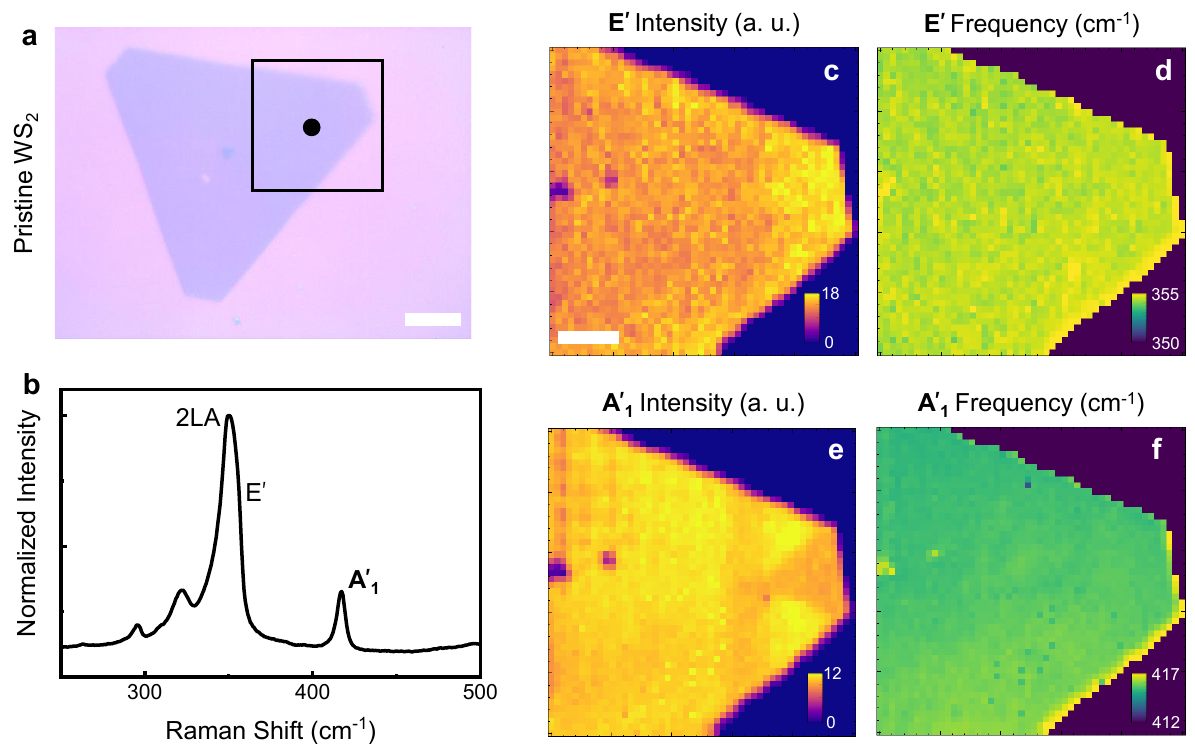} 
	\caption{{\bf Raman hyperspectral mapping of a pristine WS$_2$ monolayer.}
    \textbf{a,} Optical image of a CVD-grown pristine WS$_2$ monolayer. Scale bar: 10~$\mu$m.
    \textbf{b,} Representative Raman spectrum acquired from the position indicated by the black dot in ({\bf a}). The spectrum displays the characteristic second-order 2LA mode and the first-order $E^{\prime}$ and $A^{\prime}_1$ vibrational modes.
    \textbf{c-f,} Hyperspectral intensity (\textbf{c,e}) and frequency (\textbf{d,f}) maps for the $E^{\prime}$ (\textbf{c,d}) and $A^{\prime}_1$ (\textbf{e,f}) modes. The spatial uniformity of both intensity and frequency across the domain confirms the high crystalline quality and absence of localized perturbations in the undoped sample. Scale bar in (\textbf{c}): 5~$\mu$m.}
	\label{SuppFig6}
\end{figure}

To isolate the effect of vanadium incorporation on the vibrational properties of WS$_2$, Raman hyperspectral mapping was performed on a pristine (undoped) monolayer for comparison. Supplementary Figure~\ref{SuppFig6}a shows an optical image of the pristine sample, with a representative Raman spectrum provided in Supplementary Fig.~\ref{SuppFig6}b. The spectrum exhibits the characteristic second-order longitudinal acoustic mode (2LA), along with the first-order in-plane $E^{\prime}$ and out-of-plane $A^{\prime}_1$ modes.

The corresponding intensity (Supplementary Figs.~\ref{SuppFig6}c,e) and frequency maps (Supplementary Figs.~\ref{SuppFig6}d,f) for the $E^{\prime}$ and $A^{\prime}_1$ modes reveal a highly uniform Raman response across the entire domain. This homogeneity stands in stark contrast to the highly V-doped samples, where the bisector lines exhibit significant frequency shifts and intensity modulations. These findings confirm that the observed spatial variations in the doped samples are directly linked to vanadium-induced perturbations of the crystal lattice and the resulting modulation of the electron-phonon coupling.

\begin{figure}[!htbp]
	\centering
	\includegraphics[width=0.75\linewidth]{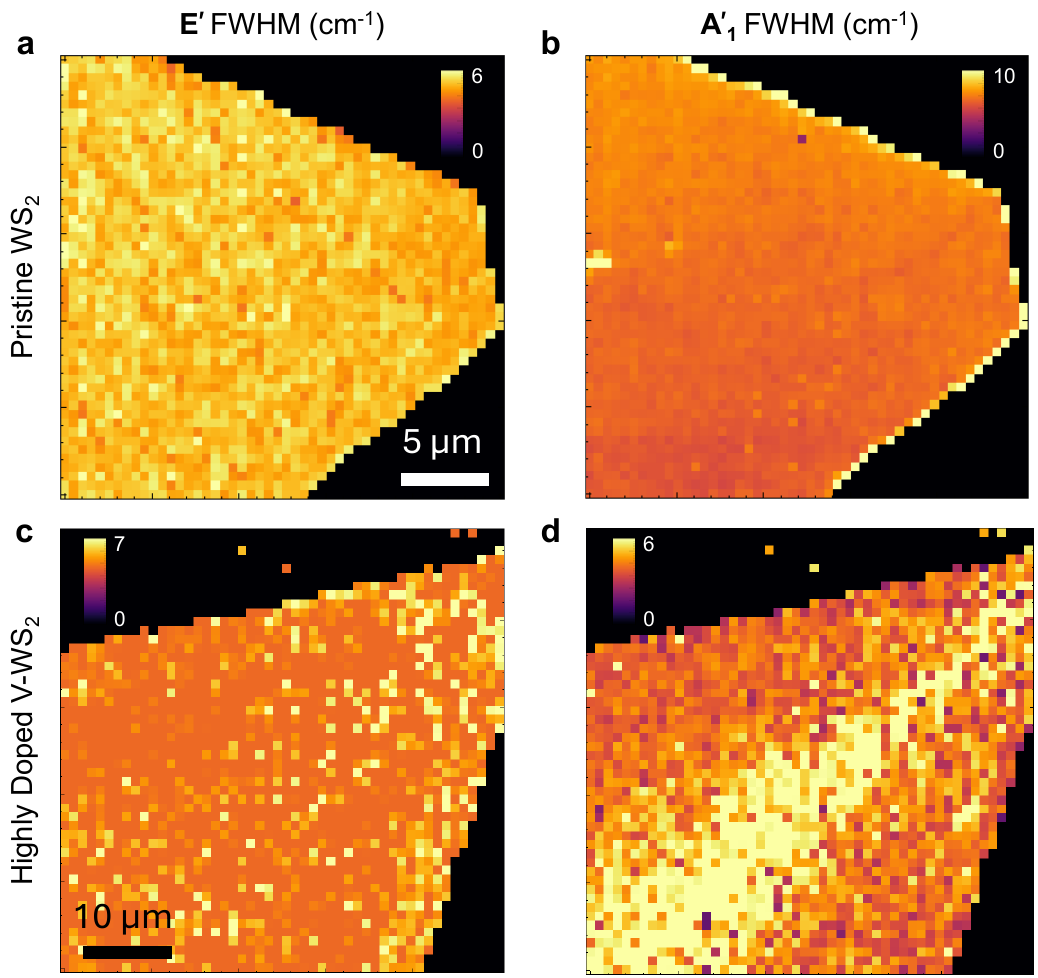} 
	\caption{{Mode-selective linewidth broadening induced by V-doping.}
    Full-width at half maximum (FWHM) Raman maps for the $E^{\prime}$ (\textbf{a,c}) and $A^{\prime}_1$ (\textbf{b,d}) modes. \textbf{a,b,} Pristine WS$_2$ monolayer exhibiting a homogeneous linewidth distribution for both in-plane and out-of-plane modes.
    \textbf{c,d,} Highly V-doped WS$_2$ monolayer. While the $E^{\prime}$ mode ({\bf c}) remains relatively uniform, the $A^{\prime}_1$ mode ({\bf d}) shows significant localized broadening along the bisector trajectories. Scale bars: 5~$\mu$m ({\bf a}) and 10~$\mu$m ({\bf c}).}
	\label{SuppFig7}
\end{figure}

Supplementary Figure~\ref{SuppFig7} provides the full-width at half maximum (FWHM) maps for the $E^{\prime}$ and $A^{\prime}_1$ modes in both pristine and highly V-doped monolayers. While the pristine sample maintains a uniform linewidth for both modes (Supplementary Figs.~\ref{SuppFig7}a,b), the highly doped sample displays a distinct, mode-selective broadening. Specifically, the $A^{\prime}_1$ mode undergoes significant broadening along the bisector trajectory (Supplementary Fig.~\ref{SuppFig7}d), whereas the $E^{\prime}$ mode remains relatively uniform (Supplementary Fig.~\ref{SuppFig7}c). This localized broadening of the out-of-plane $A^{\prime}_1$ mode is a characteristic signature of symmetry-breaking disorder and phonon scattering induced by a high concentration of substitutional vanadium atoms~\citeB{zou2023raman}.

\begin{figure}[!htbp]
	\centering
	\includegraphics[width=0.75\linewidth]{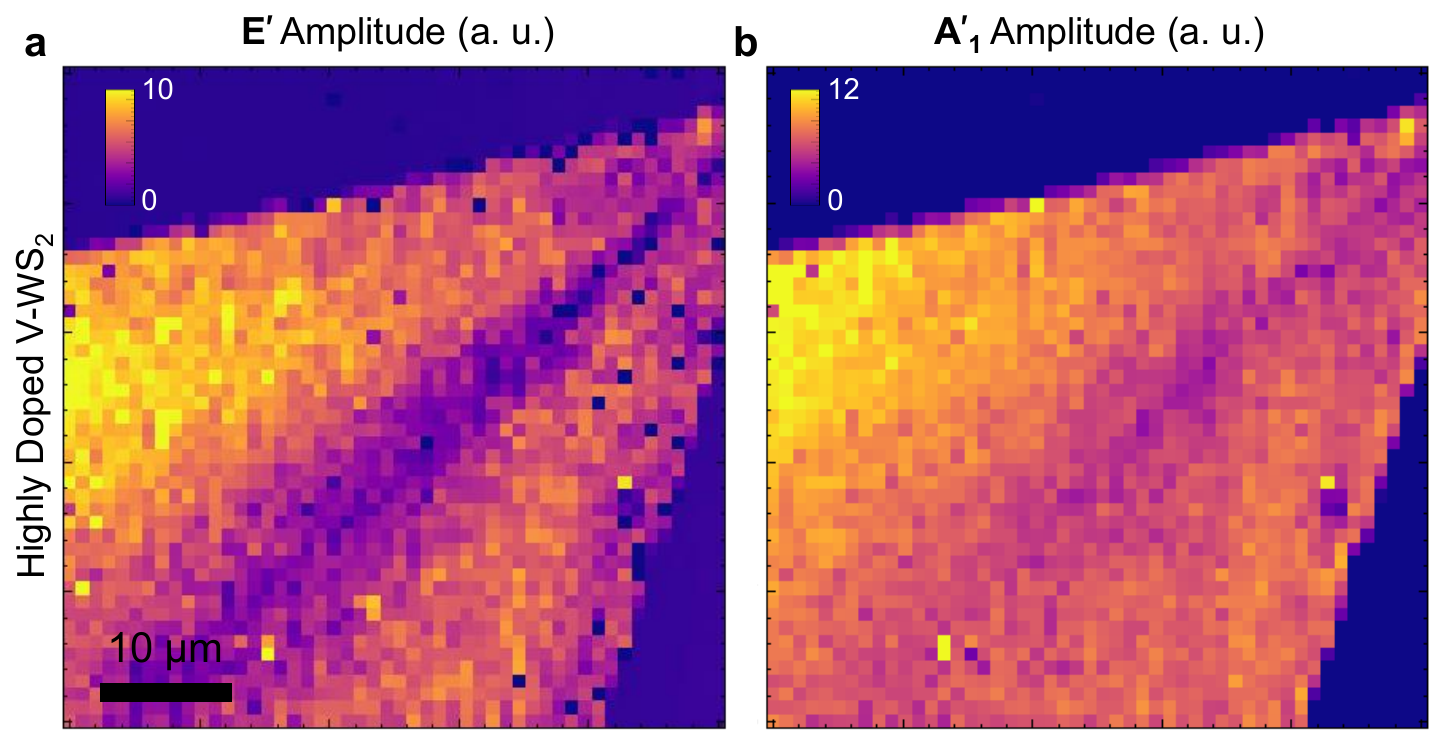} 
	\caption{{Mode-selective amplitude decay induced by V-doping.}
    Raman amplitude maps for the $E^{\prime}$ (\textbf{a}) and $A^{\prime}_1$ (\textbf{b}) modes in the highly V-doped WS$_2$ monolayer. 
    \textbf{a,} $E^{\prime}$ amplitude exhibiting an amplitude quenching at the bisector line, following the integrated intensity response.
    \textbf{b,} $A^{\prime}_1$ amplitude presenting a slight decrease at the bisector line, which accounts for its uniform integrated intensity despite the linewidth broadening. 
    Scale bar: 10~$\mu$m ({\bf a}).}
	\label{SuppFig8}
\end{figure}

Supplementary Figure~\ref{SuppFig8} shows the $E^{\prime}$ and $A^{\prime}_1$ amplitude maps for the highly-doped WS$_2$ monolayer. The in-plane $E^{\prime}$ mode exhibits a strong reduction in intensity at the bisector line, in accordance with the integrated area behavior reported in Fig.~3c. The out-of-plane $A^{\prime}_1$ mode also displays an amplitude decrease at the bisectors, contrasting with the nearly uniform integrated intensity response shown in Fig.~3e. As discussed in the main text, this is explained by the $A^{\prime}_1$ linewidth broadening in the segregated region, which compensates the reduced intensity and leads to a homogeneous area. Nevertheless, even for the amplitude maps, we observe a significantly higher modulation for the $E^{\prime}$ mode, which corroborates the stronger impact of alloying in the in-plane vibrations.

%%%%%%%%%%%%%%%%%%%%%%
\clearpage
\section*{\large{Supplementary Note 7: Topographic analysis}}

The Raman spectra analysis discussed in the main text revealed an unexpected tensile strain localized at the bisector lines of the V-doped WS$_2$ monolayers. To better understand the origin of this structural response, we performed a comprehensive topographic analysis using both atomic force microscopy (AFM) and optical contrast profiling.

\begin{figure}[!htbp]
	\centering
	\includegraphics[width=1.0\linewidth]{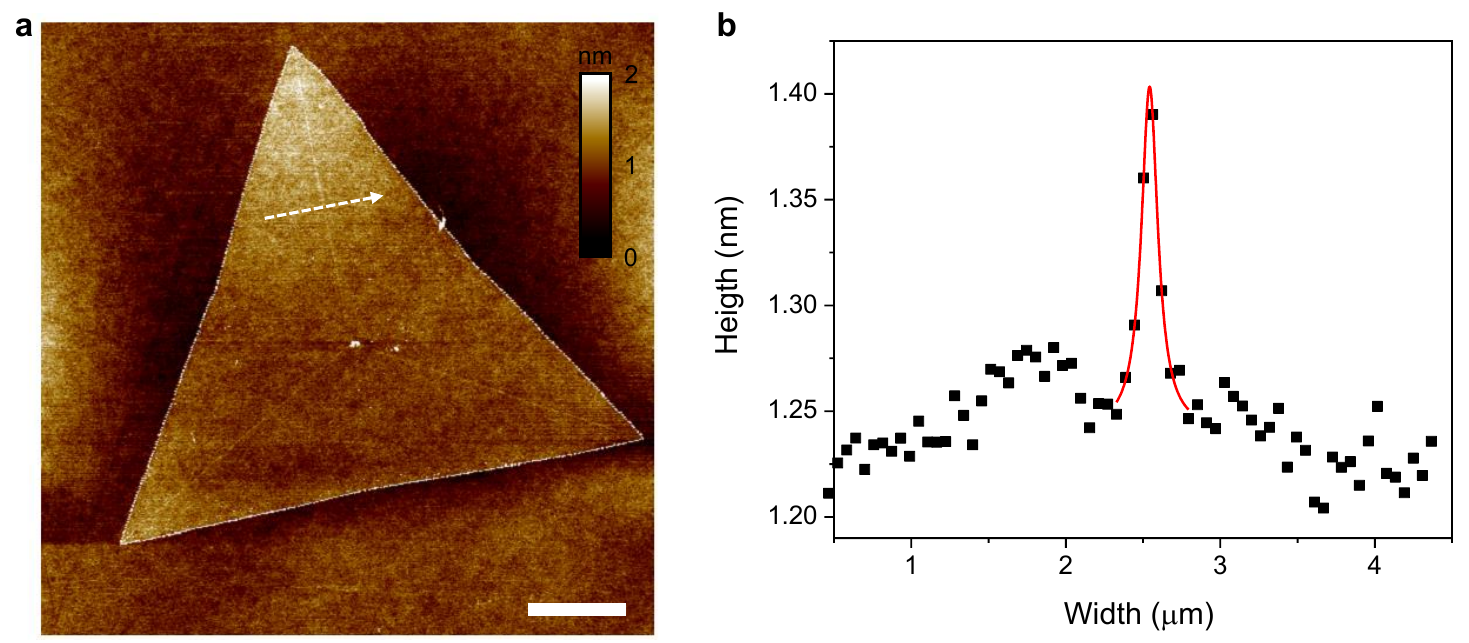} 
	\caption{{\bf Topographic uplift in moderately V-doped WS$_2$.}
    \textbf{a,} Atomic force microscopy (AFM) height map of a moderately doped domain. The bisector lines exhibit a clear vertical elevation relative to the domain interior. Scale bar: 5~$\mu$m.
    \textbf{b,} Representative height line profile (taken across the white dashed line) showing a maximum uplift of 1.2~nm and a lateral width of 130~nm, providing the structural basis for the tensile strain.}
	\label{SuppFig9}
\end{figure}

For the moderately doped sample, AFM experiments were conducted on freshly grown samples (Supplementary Fig.~\ref{SuppFig9}a). The measurement reveals a distinct topographic ``uplift'' or wrinkling localized along the bisector lines. This morphological elevation explains the reported tensile strain, as the vertical displacement relative to the substrate induces a lateral stretching of the WS$_2$ lattice. The height line profile displayed in the Supplementary Fig.~\ref{SuppFig9}b reveals a maximum elevation of 1.2~nm at the bisector, with full-width at half maximum (FHWM) of 130~nm. This confirms the highly localized nature of the mechanical deformation. Our observations of topographic uplift in these moderately doped samples are consistent with previous reports for both triangular- and hexagonal-shaped V- and Fe-doped WS$_2$ monolayers~\citeB{zhang2023spatial,menescal2026resonance}, highlighting the reproducibility of this doping-dependent topographic response.

Regarding the highly V-doped WS$_2$ monolayers, scanning probe measurements were conducted after the completion of XRF and Raman hyperspectral mapping reported in this study. While the samples remained suitable for optical and X-ray probes, scanning probe techniques like AFM are significantly more sensitive to surface contaminants and adsorbates that accumulate over time. Consequently, the topographic maps for the highly doped samples were strongly affected by surface roughness, precluding a quantitative AFM height determination. To address this limitation, we employed an established methodology for qualitatively assessing sample topography via optical contrast ($OC$) analysis~\citeB{castellanos2010optical,li2012optical}.

\begin{figure}[htb]
	\centering
	\includegraphics[width=0.8\linewidth]{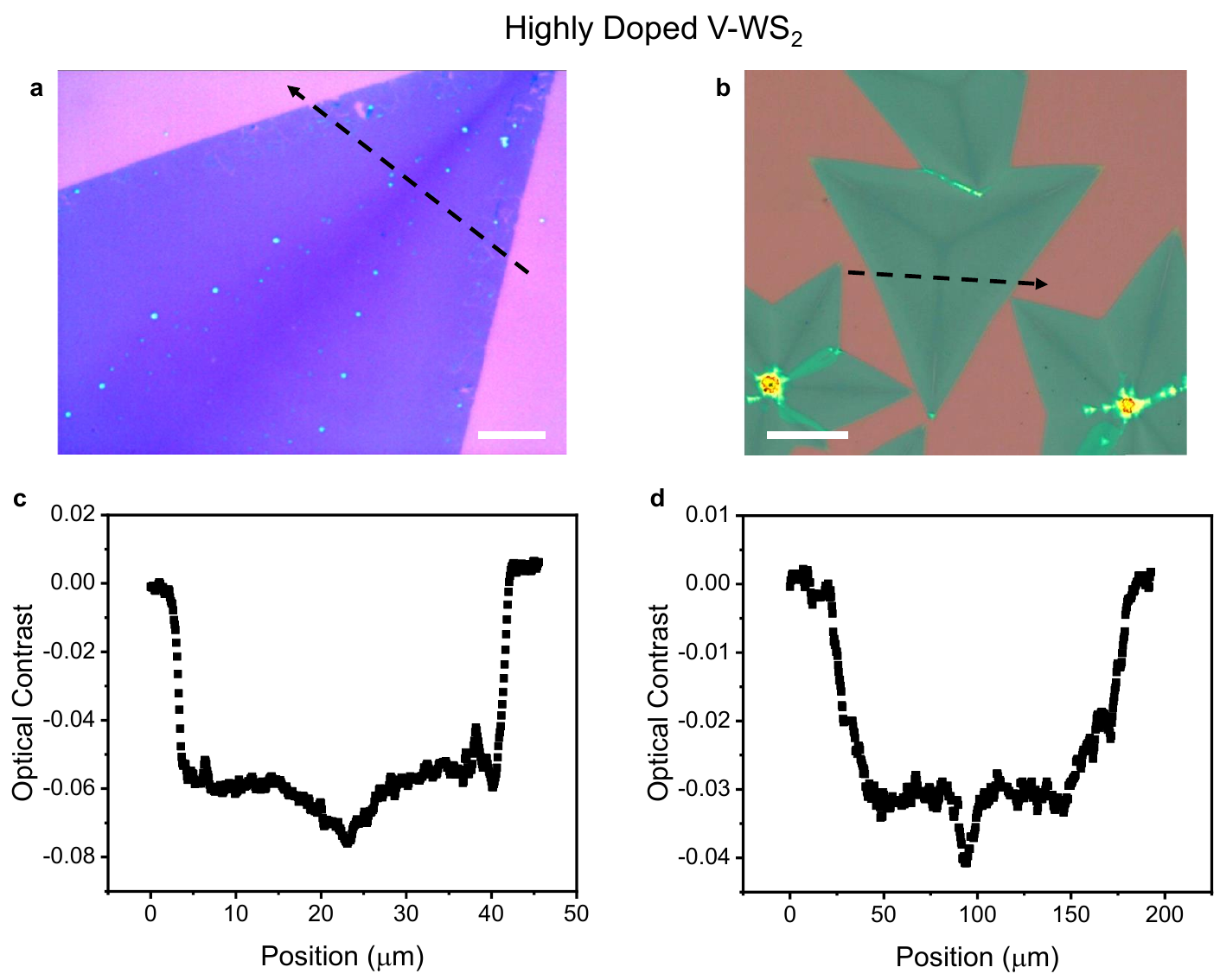} 
	\caption{{\bf Optical contrast analysis of highly V-doped WS$_2$.}
    \textbf{a,b,} Optical images of highly doped monolayers. Scale bars: 10~$\mu$m ({\bf a}) and 100~$\mu$m ({\bf b}).
    \textbf{c,d,} Optical contrast ($OC$) line profiles extracted along the black dashed arrows in ({\bf a}) and ({\bf b}). The localized contrast reduction at the bisector lines qualitatively confirms the persistence of topographic uplift in the high-doping regime.}
	\label{SuppFig10}
\end{figure}

Supplementary Figures~\ref{SuppFig10}a,b show optical images of the highly doped monolayers used in the present study. Both monolayers exhibit a distinct dark contrast localized specifically at the bisector lines. We quantified this feature by calculating the optical contrast according to~\citeB{castellanos2010optical,li2012optical}:

\begin{equation}
    OC = \frac{I_{flake} - I_{substrate}}{I_{flake} + I_{substrate}},
\end{equation}

\noindent where $I_{flake}$ and $I_{substrate}$ represent the intensities of the flake and substrate, respectively. The resulting profiles (Supplementary Figs.~\ref{SuppFig10}c,d) confirm a significant modulation of the optical contrast along the bisectors. Following established optical models for 2D materials on SiO$_2$/Si, this contrast variation serves as a robust qualitative indicator of an elevated topography, corroborating the presence of mechanical wrinkling/uplift even in the high-doping limit.

%%%%%%%%%%%%%%%%%%%%%%

\newpage
\section*{\large{Supplementary Note 8: Vibrational calculations}}

Within the 2H structural regime of V-doped WS$_2$ monolayers, the vibrational response separates naturally into two distinct classes. The first class comprises host-derived phonons, including the $E'$- and $A_1'$-like modes, whose frequencies respond primarily to the overall mechanical state of the monolayer. The second class consists of impurity-activated modes that arise from the local chemical and structural environment of the V dopants and their surrounding sulfur coordination shells. This distinction is essential for interpreting the bisector spectra, because the experiment simultaneously probes a mesoscale tensile field and a locally perturbed impurity environment.

\begin{figure}[ht]
    \centering
    \includegraphics[width=0.75\linewidth]{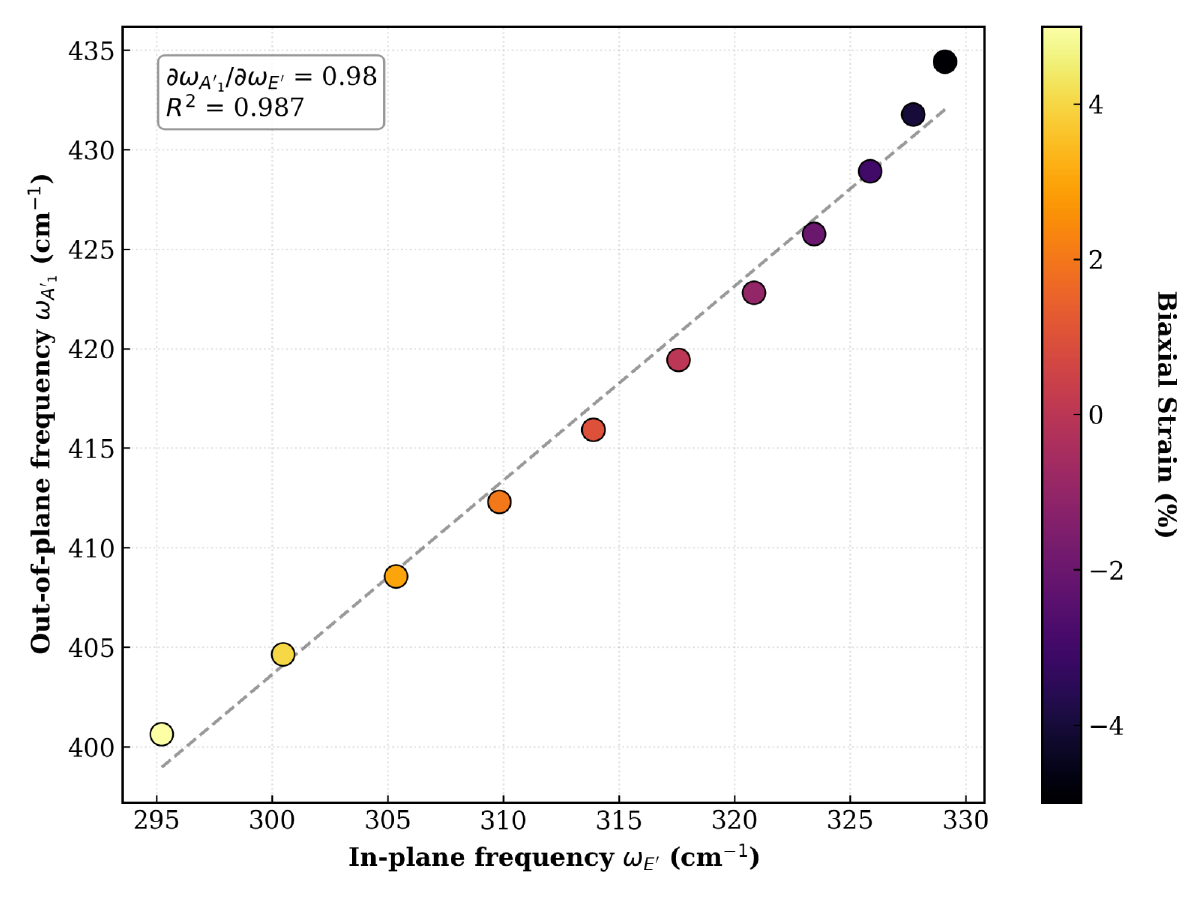}
    \caption{\textbf{Theoretical strain correlation between the host Raman-active modes in WS$_2$ monolayers.}
    DFT-calculated frequency correlation between the in-plane E$'$ mode and the out-of-plane A$_1'$ mode under systematically varied biaxial strain. The dashed line represents the best-fit mechanically driven trend, with slope $\partial\omega_{A_1'}/\partial\omega_{E'} = 0.98$ and $R^2 = 0.987$. The color scale indicates the applied biaxial strain, ranging from tensile to compressive conditions. This manifold is used here as a theoretical reference for the mechanically induced evolution of the host phonons. Comparison with the experimental bisector shifts supports the presence of a localized tensile field of approximately $0.70\%$, while deviations from the ideal line reflect the additional influence of composition, local defect chemistry, and defect-assisted electronic renormalization.}
    \label{fig:strain_correlation}
\end{figure}

To quantify the mechanical contribution to host-phonon renormalization, we calculated the strain-dependent correlation between the in-plane $E'$ and out-of-plane $A'_1$ modes of pristine WS$_2$ under systematically varied isotropic biaxial strain. As shown in Fig.~\ref{fig:strain_correlation}, the DFT data define a nearly linear mechanically driven trajectory with slope $\partial \omega_{A_1'}/\partial \omega_{E'} = 0.98$ ($R^2 = 0.987$). We stress that this manifold should be regarded as a theoretical reference for the mechanically induced trend, rather than as a universal calibration curve. Even so, its comparison with the experimental bisector shifts provides a useful consistency check and agrees with a localized tensile strain of approximately $0.70\%$. This value is fully compatible with the experimentally observed topographic uplift and with the softening of the host modes, indicating that the bisector response is governed predominantly by structural relaxation rather than by intrinsic chemical stiffening alone.

\begin{figure}[!htb]
    \centering
    \includegraphics[width=0.8\linewidth]{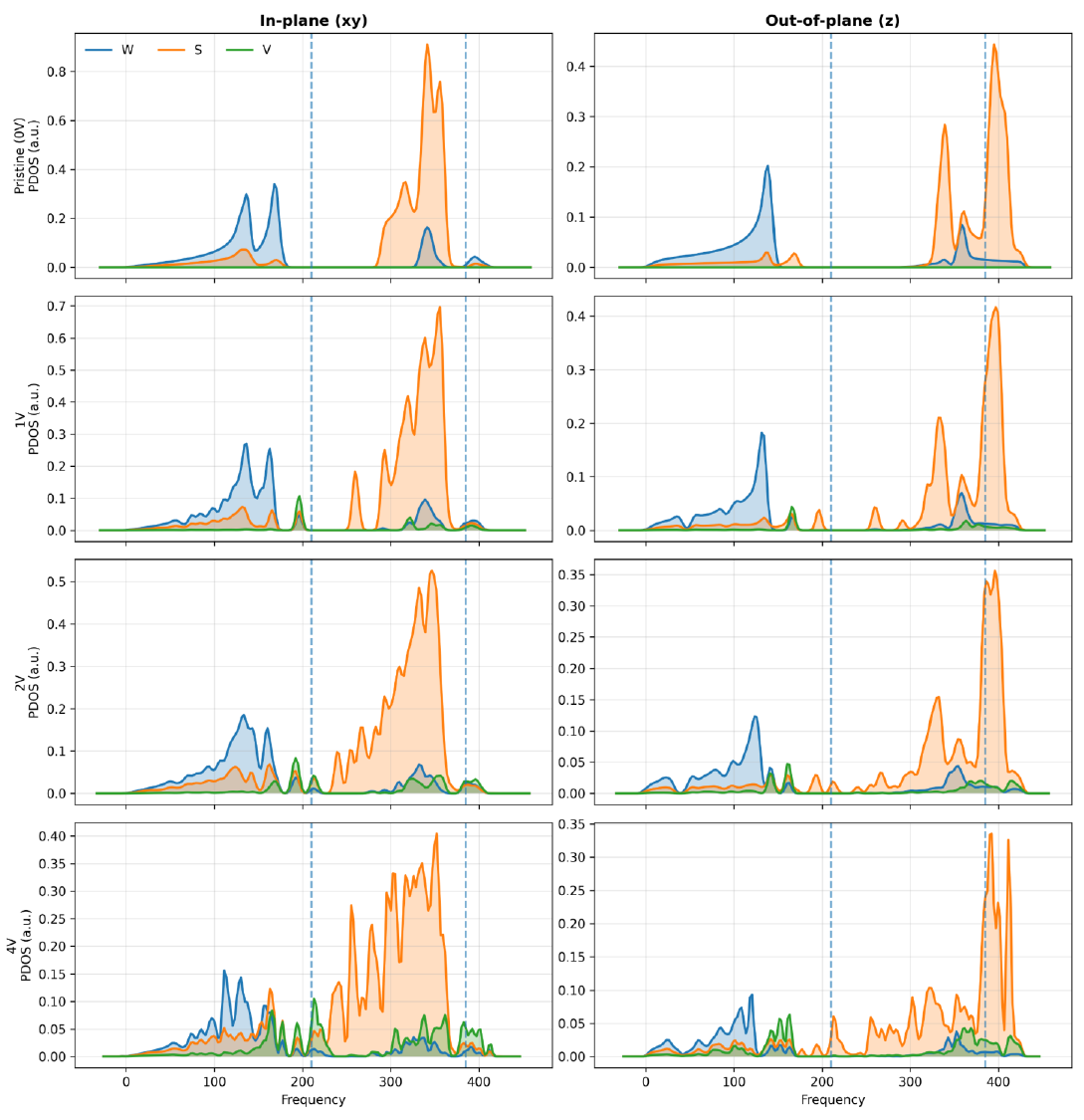}
    \caption{{\bf Species-resolved vibrational density of states in V$_x$W$_{1-x}$S$_2$ monolayers.} 
    Atomic decomposition of the Phonon Density of States (PDOS) into in-plane ($xy$, left column) and out-of-plane ($z$, right column) displacements. The vibrational spectra are partitioned into specific contributions from Tungsten (W, blue), Sulfur (S, orange), and Vanadium (V, green) for concentrations ranging from the pristine lattice (0V) to the 4V configuration. Vertical dashed lines at $\sim 210$ and $\sim 385$ cm$^{-1}$ indicate the emergence of $V$-induced impurity bands, corresponding to the $J2$ and $J3$ modes.}
    \label{fig:vibrational_species}
\end{figure}

To gain deeper insight into the impurity-activated modes, the species-resolved phonon density of states (PDOS) was calculated and decomposed into in-plane ($xy$) and out-of-plane ($z$) vibrational components for the pristine, 1V, 2V, and 4V configurations (Figure~\ref{fig:vibrational_species}). Two impurity-related spectral windows emerge systematically upon V incorporation: one near $\sim 210$ cm$^{-1}$ and a second near $\sim 385$ cm$^{-1}$. The lower-energy feature exhibits strong V-derived in-plane character, whereas the higher-energy feature is dominated by sulfur motion with a more pronounced out-of-plane component. These trends already suggest a natural correspondence with the experimental $J2$ and $J3$ signatures.

The calculated Raman spectra in Fig.~4c support this interpretation. In the ordered 2V supercell, the most prominent defect-related mode appears at $216.9$ cm$^{-1}$ and is dominated by in-plane V motion, with a clear counterpart in the $xy$-projected PDOS. We therefore assign this feature as the most plausible microscopic origin of the experimental $J2$-like band. In the dilute 1V limit, its precursor already appears at $197.1$ cm$^{-1}$, indicating that the mode is activated even for isolated substitutional impurities but becomes substantially more prominent as the local symmetry breaking and impurity-induced distortions increase. 

%\begin{figure}[!htb]
%    \centering
%    \includegraphics[width=0.75\linewidth]{SuppFigures/SFig11.pdf}
%    \caption{{\bf Calculated Raman response of vanadium-doped WS$_2$ monolayers.} Simulated Raman spectra for pristine, 1V, and 2V ($3 \times 3 \times 1$) configurations computed within the Placzek approximation. The calculations employ a dense $10 \times 10 \times 1$ k-point grid to ensure convergence of the dielectric tensor derivatives ($\partial \varepsilon / \partial Q$). The vertical dashed lines indicate the positions of the primary $E'$ and $A'_1$ modes, alongside the emergent features corresponding to the experimental $J2$ and $J3$ signatures. The progressive broadening and activation of additional modes in the 1V and 2V cases illustrate the impact of dopant-induced symmetry breaking and local structural frustration on the phononic landscape.}
%    \label{fig:raman_calc}
%\end{figure}

The corresponding feature near $387.5$--$387.9$ cm$^{-1}$ is best described as a $J3$-like mode associated primarily with the sulfur environment of the impurity. In the 1V configuration, this vibration is strongly localized on the first sulfur coordination shell around the V atom, whereas in the 2V case, it evolves into a more collective sulfur-dominated excitation with mixed metal participation.

The lower-frequency portion of the spectrum further clarifies how the dopant perturbs the lattice. In the 1V configuration, the $149.1$ cm$^{-1}$ mode corresponds to a breathing-like motion of the surrounding W network around an almost stationary V center, indicating that the impurity acts as a local pinning site for the host metal framework. The modes at $288.6$ and $322.8$ cm$^{-1}$ probe the immediate V$-$S environment more directly: the former is dominated by the first-neighbor sulfur coordination shell and can be viewed as a localized breathing--stretching vibration, while the latter involves coupled in-plane motion of V and neighboring S atoms and therefore reflects the local chemical anisotropy introduced by the substitutional defect. Their 2V counterparts preserve the same qualitative character but become more collective as neighboring impurity perturbations begin to interact.

\begin{figure}[!htb]
    \centering
    \includegraphics[width=.58\linewidth]{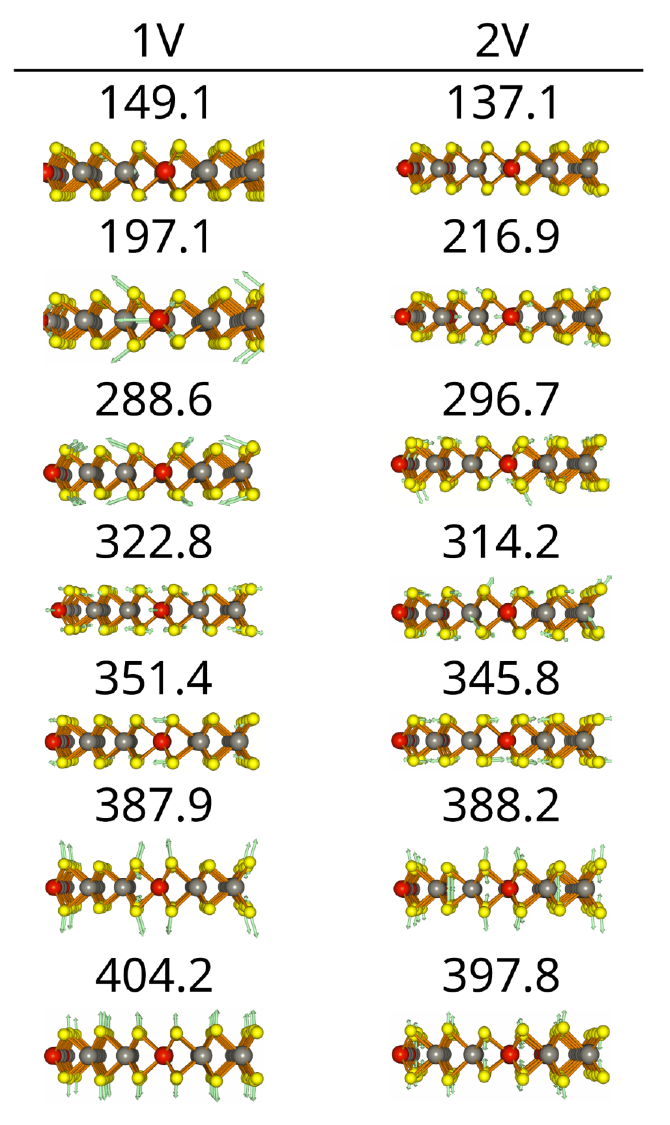}
    \caption{\textbf{Atomic displacement patterns of representative Raman-active modes in the 1V and 2V supercells.}
    Side-view snapshots of the phonon eigenvectors for the most representative Raman-active vibrations listed in Table~\ref{tab:raman_10x10x1_comparison}. Frequencies are indicated in cm$^{-1}$ above each panel. The modes near 197.1 and 216.9 cm$^{-1}$ correspond to the J$_2$-like impurity vibrations and are dominated by in-plane vanadium motion. The features near 387.9 and 388.2 cm$^{-1}$ correspond to J$_3$-like modes and involve strong sulfur participation in the immediate impurity environment. The lower-energy mode at 149.1 cm$^{-1}$ illustrates the breathing-like response of the W host network around a nearly pinned V center, whereas the intermediate-frequency modes near 288.6/296.7 and 322.8/314.2 cm$^{-1}$ highlight localized and collective vibrations of the V--S coordination shell. The highest-frequency modes near 404.2 and 397.8 cm$^{-1}$ retain predominantly host-like A$_1'$ character but already reflect the symmetry lowering induced by V incorporation.}
    \label{fig:vibrational_modes}
\end{figure}

These assignments are summarized in Table~\ref{tab:raman_10x10x1_comparison}, while the corresponding phonon eigenvectors are shown in Fig.~\ref{fig:vibrational_modes}. Importantly, the impurity-induced J$_2$- and J$_3$-like features should be distinguished from the redshift of the host $E'$ and $A_1'$ modes. The former are controlled mainly by the local defect geometry and the immediate V$-$S coordination shell, whereas the latter are sensitive to the broader mechanical state of the monolayer.

\begin{table*}[ht]
\centering
\caption{\textbf{Representative impurity-activated and host-derived Raman-active modes in V$_x$W$_{1-x}$S$_2$ monolayers computed with a $10\times10\times1$ \textit{k}-point mesh.} For each composition, the listed mode corresponds to the most Raman-active phonon within the indicated spectral window. Because absolute Raman activities can strongly vary across structurally distinct supercells, the discussion emphasizes mode character and relative prominence within each composition.}
\label{tab:raman_10x10x1_comparison}

\scriptsize
\setlength{\tabcolsep}{4pt}
\renewcommand{\arraystretch}{1.15}

\begin{tabular}{p{1.cm} p{2.25cm} p{0.8cm} p{1.7cm} p{1.9cm} p{2.7cm} p{3.2cm}}
\toprule
\textbf{} &
\textbf{Spectral window} &
\textbf{Mode} &
\textbf{Frequency (cm$^{-1}$)} &
\textbf{Activity (\AA$^4$/amu)} &
\textbf{Dominant character} &
\textbf{Assignment} \\
\midrule

\textbf{1V} &
Low-energy defect-like &
63 &
149.121 &
$1.30\times10^{6}$ &
W$_{xy}$ around pinned V &
defect breathing \\

&
J$_2$-like &
55 &
197.060 &
$2.10\times10^{4}$ &
V$_{xy}$ &
impurity mode \\

&
280--300 cm$^{-1}$ &
52 &
288.641 &
$7.76\times10^{5}$ &
S$_{xy}$ (first shell) &
local V--S breathing--stretching \\

&
320--340 cm$^{-1}$ &
43 &
322.809 &
$3.07\times10^{5}$ &
mixed V$_{xy}$ + S$_{xy}$ &
local coupled vibration \\

&
E$'$-like &
19 &
351.437 &
$1.56\times10^{5}$ &
host-like in-plane &
softened host mode \\

&
J$_3$-like &
8 &
387.878 &
$1.55\times10^{5}$ &
S-dominated, mixed polarization &
impurity mode \\

&
A$_1'$-like &
2 &
404.179 &
$1.72\times10^{6}$ &
host-like out-of-plane &
softened host mode \\

\midrule

\textbf{2V} &
Low-energy defect-like &
66 &
137.058 &
$1.254\times10^{10}$ &
mixed metal-shell motion &
collective defect mode \\

&
J$_2$-like &
55 &
216.856 &
$1.848\times10^{11}$ &
V$_{xy}$ &
impurity mode \\

&
280--300 cm$^{-1}$ &
45 &
296.666 &
$2.438\times10^{10}$ &
S$_{xy}$ + local metal coupling &
collective local vibration \\

&
320--340 cm$^{-1}$ &
39 &
314.232 &
$1.465\times10^{11}$ &
mixed V$_{xy}$ + S$_{xy}$ &
collective defect mode \\

&
E$'$-like &
23 &
345.817 &
$4.879\times10^{10}$ &
host-like in-plane &
softened host mode \\

&
J$_3$-like &
7 &
387.535 &
$1.400\times10^{11}$ &
S-dominated, mixed polarization &
impurity mode \\

&
A$_1'$-like &
6 &
397.835 &
$3.412\times10^{8}$ &
host-like out-of-plane &
softened host mode \\

\bottomrule
\end{tabular}
\end{table*}

This distinction helps rationalize the apparently contrasting signatures observed experimentally and theoretically. Our local structural analysis shows that the impurity-associated $J2$-like vibration is governed by a compressed V$-$S coordination shell, whereas hyperspectral Raman mapping reveals that the bisectors exhibit an overall tensile strain field that softens the host phonons. Therefore, the Raman response reflects a multiscale lattice picture: local chemical compression contributes to the activation and frequency placement of the impurity modes, while extended tensile relaxation softens the host $E'$ and $A_1'$ branches. In this sense, the vibrational anomalies are not competing interpretations, but complementary manifestations of the same segregated defect architecture.

Overall, the Raman calculations converge toward a unified multiscale picture. Vanadium incorporation locally compresses and chemically reorganizes the metal-chalcogen framework, thereby activating impurity modes such as the J$_2$- and J$_3$-like bands. At the same time, when V-rich regions segregate and interact over larger length scales, the monolayer undergoes tensile mechanical relaxation that softens the host $E'$ and $A_1'$ modes. The vibrational anomalies observed at the bisectors thus emerge from the combined action of local defect chemistry and mesoscale strain accumulation.

%\begin{figure}[!htb]
%    \centering
%    \includegraphics[width=1\linewidth]{SuppFigures/SFig14.pdf}
%    \caption{\textbf{Effective Band Structure (EBS) and orbital hybridization.} \textbf{a--d}, Unfolded band structures of the $3 \times 3$ WS$_2$ supercell into the primitive Brillouin zone using the EBS method for: (\textbf{a}) substitutional 1V doping, (\textbf{b}) isolated $V_S$, (\textbf{c}) co-doped 1V+$V_S$ complex, and (\textbf{d}) high-concentration 3V+$V_S$ regime. \textbf{e,f}, Orbital-projected EBS for cases (\textbf{c}) and (\textbf{d}), highlighting contributions from W (blue), S (red), and V (green). The spectral weight reveals the impact of defects and the emergence of a dispersive impurity band in (\textbf{d}).}
%    \label{fig:ebs}
%\end{figure}

%%%%%%%%%%%%%%%%%%%%%%

\clearpage
\newpage
\section*{\large{Supplementary Note 9: Doping engineering via temporal flux modulation}}

To demonstrate the utility of the AGD model as a predictive design tool for growth protocol, we propose a synthesis protocol aimed at engineering complex quantum systems. By exploiting the relationship between temporal precursor flux and spatial dopant distribution, the model can be used to 'program' the position of quantum confinement zones within the monolayer.

\begin{figure}[!htbp]
	\centering
	\includegraphics[width=0.9\linewidth]{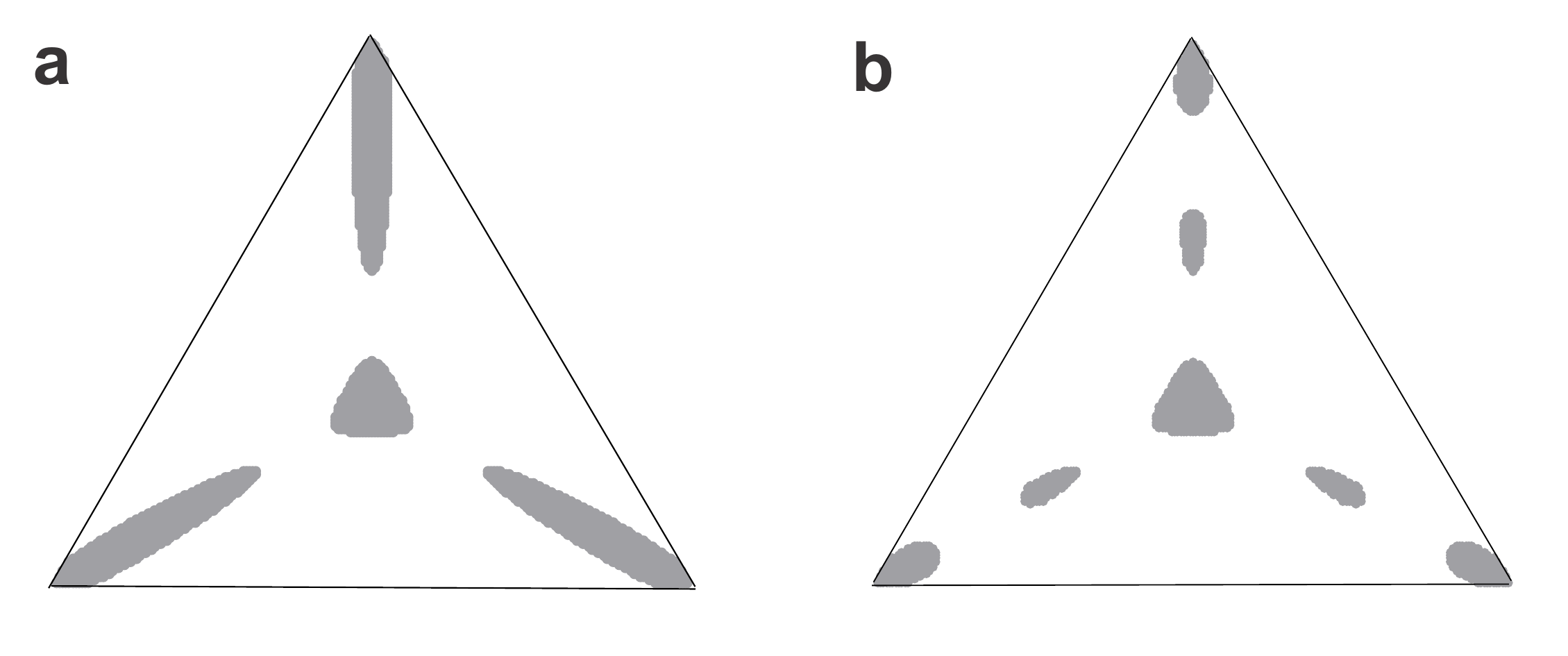} 
	\caption{{\bf Deterministic doping engineering via flux modulation.}
    Simulations highlighting regions with dopant concentrations exceeding, by more than 5$\%$, the concentration at the saddle point along the bisector lines.
    {\bf a,} Prediction for the continuous growth regime, showing an isolated central quantum dot and three vertex contacts.
    {\bf b,} Proposed architecture for growth protocol resulting from a programmed interruption of the vanadium flux (off at 60\%, on at 80\%). The modulations result in a complex architecture containing a central dot, three satellite dots, and three vertex contacts.
    }
	\label{SuppFig15}
\end{figure}

Supplementary Figure~\ref{SuppFig15}a shows the simulated distribution for the continuous growth process. By isolating regions where the dopant concentration exceeds the local saddle point concentration by more than 5$\%$, the model predicts the formation of a deterministic central quantum dot and three vertex-adjacent zones. This architecture suggests an integrated system consisting of a central 0D confinement site and three terminal contacts within the monolayer.

As proof-of-concept for more advanced defect engineering, we propose a modified growth sequence (Supplementary Fig.~\ref{SuppFig15}b) where the vanadium flux is interrupted at 60\% of the total growth time and resumed at 80\%. This proposed modulation results in a significantly more complex phenomenology. While the central quantum dot remains preserved, three additional satellite-like dots emerge between the center and the vertices. This demonstrates the potential of using growth kinetics to deterministically design multi-dot architectures and coupled quantum systems without the need for post-growth processing or lithography.

% -------------------- References --------------------


\end{document}